\documentclass[11pt]{article}

\usepackage[utf8]{inputenc}
\usepackage[OT1]{fontenc}

\usepackage{graphicx}

\usepackage{amsmath, amsthm, amssymb}
\theoremstyle{remark}
\newtheorem{remark}{Remark}

\usepackage{fullpage}
\usepackage{setspace}
\usepackage{natbib}

\usepackage{authblk}

\title{Modelling intensities of order flows in a limit order book}

\author[1]{Ioane Muni Toke}
\author[2]{Nakahiro Yoshida}
\affil[1]{University of New Caledonia, Noumea, New Caledonia.}
\affil[1]{Laboratoire MICS, Chair Of Quantitative Finance, CentraleSupelec, Paris, France.}
\affil[2]{Graduate School of Mathematical Sciences, University of Tokyo, Tokyo, Japan.}
\affil[1,2]{CREST, Japan Science and Technology Agency, Japan.}

\begin{document}

\maketitle

\begin{abstract}
We propose a parametric model for the simulation of limit order books.
We assume that limit orders, market orders and cancellations are submitted according to point processes with state-dependent intensities.
We propose new functional forms for these intensities, as well as new models for the placement of limit orders and cancellations. For cancellations, we introduce the concept of "priority index" to describe the selection of orders to be cancelled in the order book.
Parameters of the model are estimated using likelihood maximization.
We illustrate the performance of the model by providing extensive simulation results, with a comparison to empirical data and a standard Poisson reference.
\end{abstract}

\section{Introduction}

The limit order book is the central structure aggregating the orders of all traders to buy and sell shares of a given stock on an exchange.
It is standard to simplify the complex diversity of financial messages into three types of orders : limit orders are submitted with a (limit) price into the order book, where they wait to be matched by a counterpart for a transaction ; market orders are submitted without any price and are executed immediately ; cancellations of pending limit orders is possible at any time.
The order book can thus be viewed as a complex dual queueing system with price and time priority rules (see \cite{Abergel2015} for an introductory book treatment).

A partial theoretical treatment of this complex random system is possible under very simplistic assumptions, essentially assuming that the submission of limit orders, market orders and cancellations are basic Poisson processes \citep{Cont2010, MuniToke2015}.
Exact analytical results are however limited. With appropriate scaling techniques, some limit behaviours of this complex system can be studied, see e.g. \cite{Abergel2013} for a price diffusion process, or \cite{Cont2012} for a diffusion approximation of the volumes at the best quotes.

Another branch of study of the limit order books deals with a more statistical point of view.
\cite{Smith2003} investigates the order book structure with mean field techniques.
\cite{Mike2008} proposes an empirical model of the order book that aims at reproducing some of empirical observations usually made on financial markets. Among other contributions, they propose a Student model for the placement of limit orders and a three-variable model for the cancellation of pending limit orders.
The core of the submission mechanism in the order book remains however a Poisson process.
Recently, \cite{Huang2015} have proposed a model in which the intensities of submission of limit, market orders and cancellations depend on the volume of the first limit.
They are able to show that a queueing system with these intensities is able to reproduce some empirical features of the limit order book, such as the distribution of the first level.

In this paper we propose a general model in line with previous contributions such as \cite{Mike2008, Huang2015}.
We do not extend or specify previous models but build directly from the data. Our goal is to provide state-dependent intensities of submissions of limit and market orders that can be used for the simulation of a "realistic" limit order book.
We adopt the following modelling principle : limit and market orders intensities should depend on both dimensions of the limit order book, namely the price dimension and the volume dimension.
The spread is an obvious choice to include the price dimension in the modelling for both types of orders.
The volume of the first level is another obvious choice for market orders, while the total volume available appears to be a good candidate for the limit orders.
We define exponential forms of intensities that are convenient for two reasons: they keep the non-negativity of intensities of point processes, and they allow for a practical a maximum-likelihood estimation.
For the cancellation process, we introduce a new "priority index" as a main modelling variable, which turns out to be very efficient.
All proposed models are fitted on a database of 10 consecutive trading days (January 17th-28th, 2011) for six different liquid stocks traded on the Paris stock exchange.

The rest of the paper is organized as follows.
Section \ref{section:Data} briefly describes the data and its preparation.
Section \ref{section:MarketOrders} provides empirical insights on the intensity of submission of market orders and build a convenient parametric model.
Section \ref{section:LimitOrders} introduces a similar model for the intensities of limit orders and provide a very flexible Gaussian mixture model for the placement of limit orders, that is able to reproduce the multi-modality of the empirical distribution.
Section \ref{section:Cancellations} shows that the "priority volume", i.e. the volume standing in front of an pending orders according to time-price priority rules is a good candidate for the modelling of the "placement" of cancellations.
Finally, Section \ref{section:Simulation} provides extensive results of our model fitted on market data and simulated. The performance of the model is analysed, in particular with respect to a standard Poisson reference.

\section{Data}
\label{section:Data}

We use data extracted from the Thomson-Reuters Tick History (TRTH) database.
We randomly select 6 liquid stocks from the CAC 40 index (i.e. stocks among the highest capitalizations exchanged at the Paris Bourse) : Air Liquide (Reuters Identification Code (ric): AIRP.PA), Alstom (ALSO.PA), BNP Paribas (BNPP.PA), Bouygues (BOUY.PA), Carrefour (CARR.PA), Electricite de France (EDF.PA).
These stocks represent a wide panel of liquidity for CAC 40 index: BNPP.PA is a heavily traded stock, one of the most traded on the Paris Stock Exchange, while EDF.PA much less actively traded and is a much smaller capitalization (EDF.PA has even since been removed from the CAC 40 index on December 21st, 2015).

For each stock, two files can be extracted from the TRTH database, which are standardly called the trades file and the quotes file.
The quotes file is a sequence of snapshots of the limit order book, listing all the modifications due to the processing of orders, each modification being timestamped with a millisecond resolution. This file can be parsed to extract an (preliminary) order flow of limit orders (increase of the available liquidity on a given side at a given price) and cancel orders (decrease of available liquidity on a given side at a given price).
The trades file is then parsed and matched to the previous (preliminary) order flow to identify and convert some of the cancel orders into market orders.

For each trading day, we keep the subset of limit orders, market orders and cancellations occurring between 9:05 in the morning and 17:25 in the afternoon, i.e. we keep the whole trading day except the first five minutes of the day, following the opening auction, and the last fives minutes of the day preceding the closing auction. The data in these very active periods seems indeed of a lesser quality and not always reliable.
In order to build the model in Sections \ref{section:MarketOrders}, \ref{section:LimitOrders} and \ref{section:Cancellations}, we keep only orders occurring on the ask side of the limit order book, and we glue the ten days of order flows as in an artificial continuous sample.
In Section \ref{section:Simulation} however, we will adopt a more practical point of view and use both sides of the book but only one day of trading at a time to fit and test the model, without any glueing of consecutive trading days.

As a result of this process of data preparation, we have for each stock and each period (one trading day or glued trading days) a list of orders (the order flow) and for each order a list of variables describing the limit order book at the time of submission : spread, volume at the best quotes, total liquidity available at the ten best quotes.

Let us add a few words on the units of this data.
Prices in the order book must be integer multiples of a tick size which is fixed by the exchange. In our sample, AIRP.PA and BNPP.PA have a $0.01$ EUR ticksize, while the four other stocks have a $0.005$ EUR ticksize.
As for the volumes, they are numbers of shares. For ease of computations and presentation, all volumes are normalized by a stock-dependent quantity equal to the median of the trade size (market orders quantities) for this stock. In order to keep this volumes integers, we round the results to the smallest larger integer (ceiling). As a results, $0$ means really no share, while $1$ is a small non-zero volume.
These remarks should explain the $x$-axis scales of the graphs of the following sections.

\section{Market orders}
\label{section:MarketOrders}

Let $N^M$ be the point process  of submission of market orders in the limit order book and let $\lambda^M$ be its instantaneous intensity.
Our goal is to identify a simple parametric model for $\lambda^M$, which should be based on meaningful variables and be easy to estimate. We therefore identify two covariates to model $\lambda^M$ : the spread $S$ and the volume at the best quote $q_1$ (on the side of submission).

Let us first investigate the spread. Using common financial knowledge, one should expect specific variations of the intensity as a function. Firstly, $\lambda^M$ should be decreasing with $S$. Indeed, if a trader needs to buy a share when $S$ is equal to one tick, he cannot gain priority in the limit order book, and therefore has to submit a buy market order to be the first to buy the best quote. On the contrary, if the spread is large, it is sufficient to submit a buy limit order just above the best bid quote to be the first in line for the next sell-initiated transaction.

We compute on our samples an estimator of the spread-dependent intensity of limit orders:
\begin{equation}
	\hat{\lambda}^M(S) = \frac{N^M(S)}{T(S)},
\end{equation}
where $N^M(S)$ is the total number of market orders submitted when the spread is equal to $S$ and $T(S)$ is the total time during which the spread is equal to $S$ in the sample.
As an illustration, $\hat{\lambda}^M(S)$ is plotted in Figure \ref{figure:MarketOrders-empIntensity-BNPP} for one of the stocks of the sample (full results will be given below).
\begin{figure}
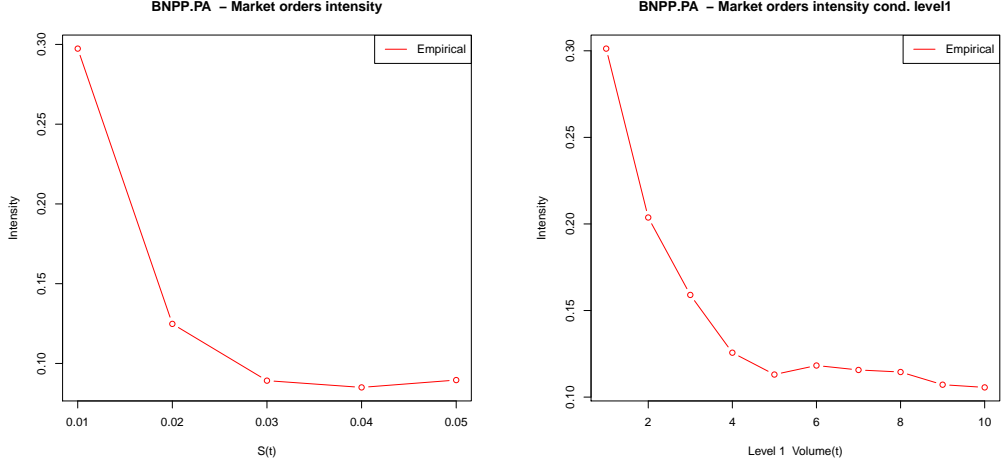

\begin{center}
\begin{tabular}{cc}
\includegraphics[width=0.4\textwidth, page=2]{{BNPP.PA-A-MarketOrders-SpreadDependence-0.99-20110117-20110128}.pdf}
&
\includegraphics[width=0.4\textwidth, page=2]{{BNPP.PA-A-MarketOrders-Level1Dependence-0.9-20110117-20110128}.pdf}
\end{tabular}
\caption{Left panel : Empirical $\lambda^M$ as a function of the spread ($\hat{\lambda}^M(S)$). Right panel : Empirical $\lambda^M$ as a function of the spread ($\hat{\lambda}^M(q_1)$). Since data may be noisy for very high values of the parameters, the $x$-axes span $99\%$ of the empirical distribution of $S$ and $90\%$ of the empirical distribution of $q_1$.}
\label{figure:MarketOrders-empIntensity-BNPP}
\end{center}
\end{figure}
As expected, the intensity of submission of market orders is decreasing with the spread.
However, this decrease does not to go to zero, and even seem to increase slightly for very large values of the spread. A plausible interpretation is that when the spread increases above usual levels, this may indicate a highly volatile period with many orders submitted. Subsequent uncertainty might translate into a "rush" for liquidity maintaining $\hat{\lambda}^M(S)$ above zero.

Following these empirical results, one might propose the following parametric model to express the functional dependence of $\lambda^M$ on $S$ :
\begin{equation}
	\lambda^M(S) = \exp \left( \beta_0 + \beta_1 \ln(S) + \beta_{11}[\ln(S)]^2 \right)
\end{equation}
The exponential form ensure that $\lambda^M$ remains non-negative. The quadratic argument allows the non-monotony of $\lambda^M$ instead of the power-law form obtain with only one term. The preference for the logarithm of the spread instead of the spread itself is detailed in Remark \ref{remark:ModelSelection}.

Let us now turn to the second explaining variable considered here, the volume $q_1$ of the best quote (on the side of submission, ask for a buy market order, bid for a sell market order).
We compute on our samples an estimator of the $q_1$-dependent intensity of limit orders:
\begin{equation}
	\hat{\lambda}^M(q_1) = \frac{N^M(q_1)}{T(q_1)},
\end{equation}
where $N^M(q_1)$ is the total number of market orders submitted when the volume on the (same side) best quote is equal to $q_1$ and $T(q_1)$ is the total time during which this volume is equal to $q_1$ in the sample. Recall that the unit for $q_1$ is the median of the trades sizes.
Results are plotted on Figure \ref{figure:MarketOrders-empIntensity-BNPP}.
One observes that $\lambda^M$ increases as $q_1$ decreases, as expected. Indeed, when $q_1$ is small, the probability that the first limit vanishes increases. This is an incentive for traders to grab the last shares available at the current price, leading to a "rush for liquidity".
This monotony however is justified for small values of $q_1$ and there is no obvious reason that the intensity should go to zero for large values of $q_1$. Figure \ref{figure:MarketOrders-empIntensity-BNPP} suggests that we can use a functional dependency on $q_1$ similar to the one suggested for the spread :
\begin{equation}
	\lambda^M(q_1) = \exp \left( \beta_0 + \beta_2 \ln(1+q_1) + \beta_{22}[\ln(1+q_1)]^2 \right)
\end{equation}

We can finally combine the two dependencies into one single model and add an potential interaction term between the two covariates. We thus obtain the following parametric model for the intensity of submission of market orders in a limit order book:
\begin{align}
	\lambda^M(t ; S(t), q_1(t)) =  \exp \bigg[
	& \beta_0 + \beta_1 \ln(S(t)) + \beta_{11}[\ln(S(t))]^2 \nonumber
	+ \beta_2 \ln(1+q_1(t)) + \beta_{22}[\ln(1+q_1(t))]^2  \nonumber
	\\ & + \beta_{12}\ln(S(t))\ln(1+q_1(t)) \bigg].
\label{equation:MarketOrders-IntensityDefinition}
\end{align}
This model can be estimated by likelihood maximization. 
To emphasize the dependency on the the parameters $\boldsymbol\beta=(\beta_0, \beta_1, \beta_{11}, \beta_2, \beta_{22}, \beta_{12})$ to be fitted, we write $\lambda^M(t ; S(t), q_1(t))=\lambda^M(t ; \boldsymbol\beta)$  when dealing with the estimation.
Log-likelihood $\mathcal L^M_T$ for the point process $\{N^M(t), t\in[0,T]\}$ as a function of the parameter vector $\boldsymbol\beta$ is defined as:
\begin{equation}
	\mathcal L^M_T(\boldsymbol\beta) = \int_0^T \ln\left(\lambda^M(t;\boldsymbol\beta)\right) \,dN^M_t - \int_0^T \lambda^M(t;\boldsymbol\beta) \,dt.
\end{equation}
Let $\{t^M_i\}$ be the set of arrival times of market orders in our sample, $\{t^S_i\}$ the set of times of jumps of the spread process $S$, $\{t^{q_1}_i\}$ the set of times of jumps of the first limit process $q_1$.
Then the log-likelihood on the sample is numerically computed as follows:
\begin{align}
	\mathcal L_T(\boldsymbol\beta) & = \beta_0 N^M(T) + \beta_1 \sum_{t^M_i} \ln S(t^M_i-) 
	+ \beta_{11} \sum_{t^M_i} [\ln S(t^M_i-)]^2 \nonumber
	\\ & + \beta_2 \sum_{t^M_i} \ln (1+q_1(t^M_i-))
	+ \beta_{22} \sum_{t^M_i} [\ln (1+q_1(t^M_i-))]^2
	+ \beta_{21} \sum_{t^M_i} \ln S(t^M_i-)\ln q_1(t^M_i-) \nonumber
	\\ & - \sum_{t_i\in\{t^S_i\}\cup\{t^{q_1}_i\} } \exp\bigg[\beta_0+\beta_1\ln S(t_i-)+\beta_{11}[\ln S(t_i-)]^2 \nonumber
	\\ & +\beta_2\ln (1+q_1(t_i-))+\beta_{22}[\ln (1+q_1(t_i-))]^2 +\beta_{12} \ln S(t_i-)\ln (1+q_1(t_i-))
	\bigg] (t_i-t_{i-1}) .
	\label{eq:MarketOrders-NumericalLogLikelihood}
\end{align}

It is then numerically maximized using the routine $\texttt{mle2}$ of the $\texttt{bbmle}$ package in the $\texttt{R}$ language. Results for all the stocks of our samples are given in Table \ref{table:MarketOrders-fittedCoeffs}. For simplicity of presentation these results are shown for the ask side only (buy market orders), but results for the bid side are similar. 
\begin{table}
\footnotesize
\begin{center}
\begin{tabular}{|l|rr|rr|rr|rr|rr|rr|}
\hline
ric & \multicolumn{1}{l}{$\beta_0$} & \multicolumn{1}{l|}{(std)} & \multicolumn{1}{l}{$\beta_1$} & \multicolumn{1}{l|}{(std)} & \multicolumn{1}{l}{$\beta_{11}$} & \multicolumn{1}{l|}{(std)} & \multicolumn{1}{l}{$\beta_2$} & \multicolumn{1}{l|}{(std)} & \multicolumn{1}{l}{$\beta_{22}$} & \multicolumn{1}{l|}{(std)} & \multicolumn{1}{l}{$\beta_{12}$} & \multicolumn{1}{l|}{(std)} \\ \hline
AIRP.PA & -0.527 & \textit{0.389} & 1.730 & \textit{0.190} & 0.370 & \textit{0.023} & -1.080 & \textit{0.091} & 0.203 & \textit{0.014} & -0.016 & \textit{0.019} \\ \hline
ALSO.PA & 6.193 & \textit{0.299} & 4.034 & \textit{0.131} & 0.537 & \textit{0.014} & -1.770 & \textit{0.061} & 0.235 & \textit{0.007} & -0.125 & \textit{0.013} \\ \hline
BNPP.PA & 3.713 & \textit{0.452} & 3.100 & \textit{0.220} & 0.482 & \textit{0.027} & -1.463 & \textit{0.058} & 0.160 & \textit{0.003} & -0.126 & \textit{0.013} \\ \hline
BOUY.PA & 1.426 & \textit{0.532} & 2.698 & \textit{0.224} & 0.425 & \textit{0.024} & -0.734 & \textit{0.098} & 0.155 & \textit{0.010} & 0.015 & \textit{0.021} \\ \hline
CARR.PA & -0.694 & \textit{0.678} & 1.565 & \textit{0.281} & 0.298 & \textit{0.029} & -0.723 & \textit{0.114} & 0.251 & \textit{0.012} & 0.111 & \textit{0.021} \\ \hline
EDF.PA & 7.863 & \textit{0.646} & 5.066 & \textit{0.269} & 0.629 & \textit{0.028} & -1.486 & \textit{0.118} & 0.154 & \textit{0.013} & -0.134 & \textit{0.022} \\ \hline
\end{tabular}
\end{center}
\caption{Maximum-likelihood parameters for the intensity $\lambda^M$ for all the stocks of our sample.}
\label{table:MarketOrders-fittedCoeffs}
\end{table}
Table \ref{table:MarketOrders-fittedCoeffs} provides the numerical values of the parameters as well as the standard deviation estimated by the maximization routine.
These standard deviations assess the quality of the fitting and verify that all the fitted values are significant to a high-level, except for small $\beta_0$'s for AIRP.PA and CARR.PA, and small $\beta_{12}$ (AIRP.PA and BOUY.PA).
This last fact concerning $\beta_{12}$ is not very surprising as the joint distribution between $S$ and $q_1$ is quite difficult to characterize, and an independence hypothesis between these two modelling variables is not unreasonable for some stocks.

\begin{figure}
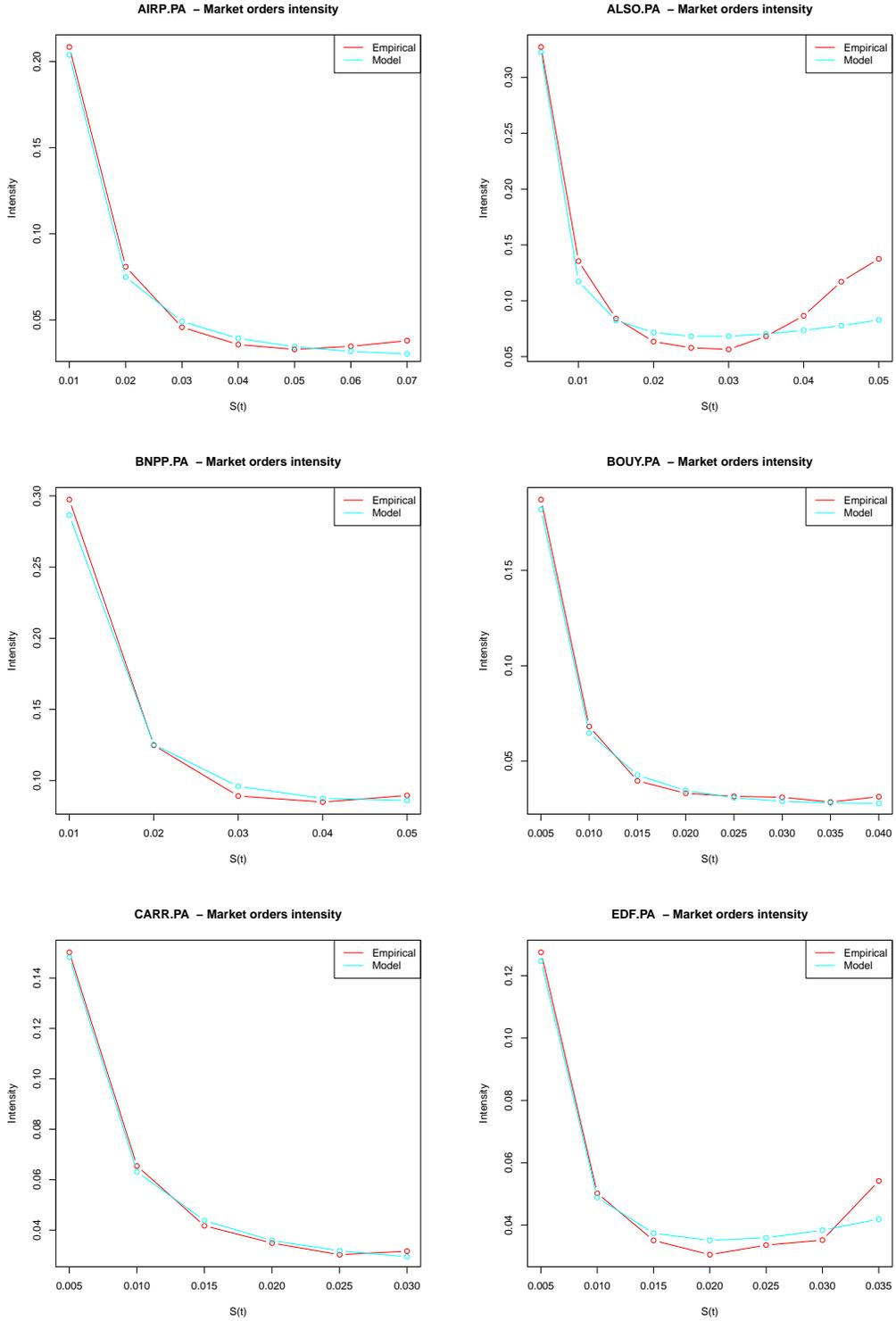

\begin{center}
\begin{tabular}{cc}
\includegraphics[width=0.4\textwidth, page=3]{{AIRP.PA-A-MarketOrders-SpreadDependence-0.99-20110117-20110128}.pdf}
&
\includegraphics[width=0.4\textwidth, page=3]{{ALSO.PA-A-MarketOrders-SpreadDependence-0.99-20110117-20110128}.pdf}
\\
\includegraphics[width=0.4\textwidth, page=3]{{BNPP.PA-A-MarketOrders-SpreadDependence-0.99-20110117-20110128}.pdf}
&
\includegraphics[width=0.4\textwidth, page=3]{{BOUY.PA-A-MarketOrders-SpreadDependence-0.99-20110117-20110128}.pdf}
\\
\includegraphics[width=0.4\textwidth, page=3]{{CARR.PA-A-MarketOrders-SpreadDependence-0.99-20110117-20110128}.pdf}
&
\includegraphics[width=0.4\textwidth, page=3]{{EDF.PA-A-MarketOrders-SpreadDependence-0.99-20110117-20110128}.pdf}
\end{tabular}
\caption{Empirical ($\hat{\lambda}^M(S)$) and model ($\tilde{\lambda}^M(S)$) intensities of market orders as functions of the spread $S$.}
\label{figure:MarketOrders-ModelIntensity-Spread}
\end{center}
\end{figure}

\begin{figure}
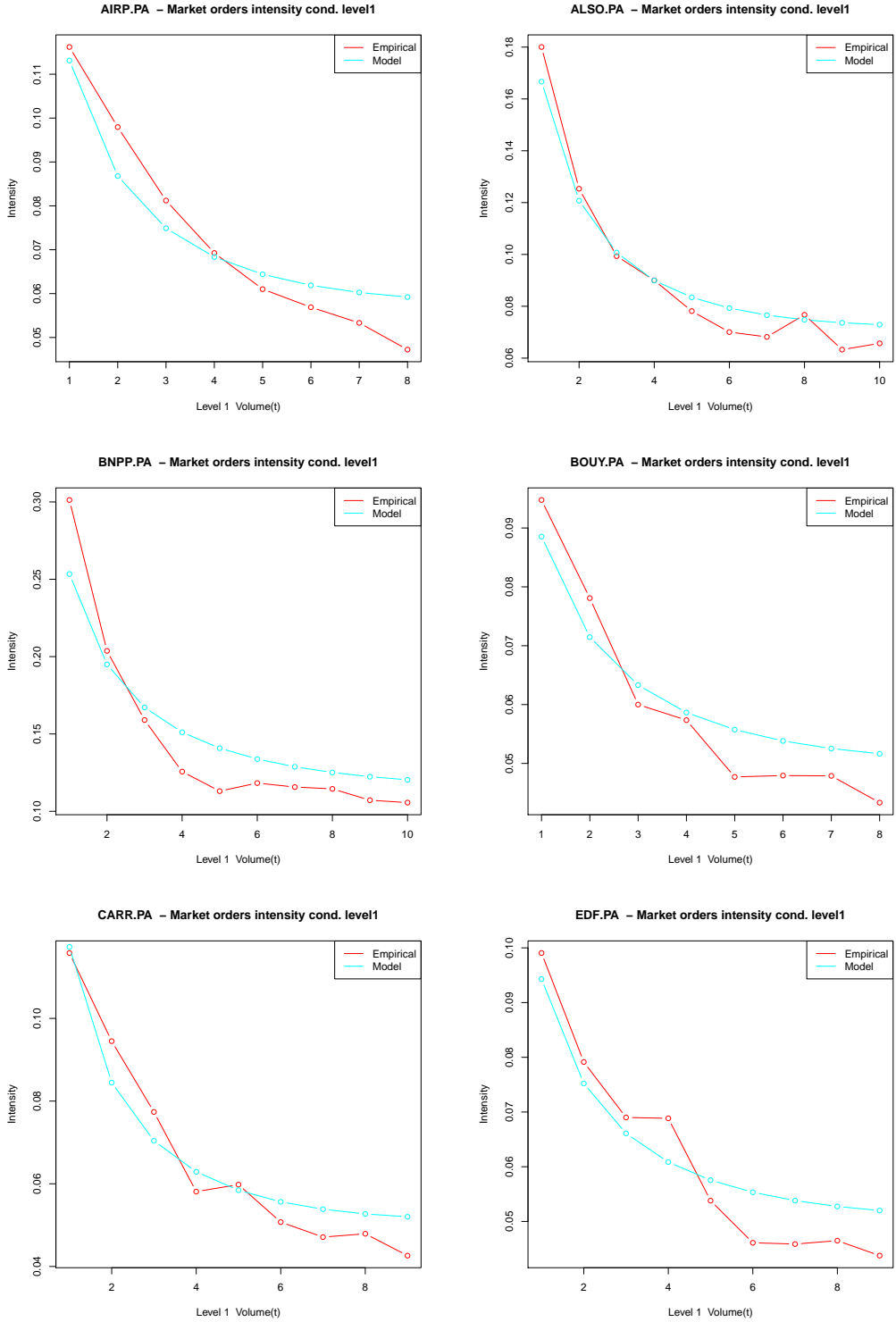

\begin{center}
\begin{tabular}{cc}
\includegraphics[width=0.4\textwidth, page=3]{{AIRP.PA-A-MarketOrders-Level1Dependence-0.9-20110117-20110128}.pdf}
&
\includegraphics[width=0.4\textwidth, page=3]{{ALSO.PA-A-MarketOrders-Level1Dependence-0.9-20110117-20110128}.pdf}
\\
\includegraphics[width=0.4\textwidth, page=3]{{BNPP.PA-A-MarketOrders-Level1Dependence-0.9-20110117-20110128}.pdf}
&
\includegraphics[width=0.4\textwidth, page=3]{{BOUY.PA-A-MarketOrders-Level1Dependence-0.9-20110117-20110128}.pdf}
\\
\includegraphics[width=0.4\textwidth, page=3]{{CARR.PA-A-MarketOrders-Level1Dependence-0.9-20110117-20110128}.pdf}
&
\includegraphics[width=0.4\textwidth, page=3]{{EDF.PA-A-MarketOrders-Level1Dependence-0.9-20110117-20110128}.pdf}
\end{tabular}
\caption{Empirical ($\hat{\lambda}^M(q_1)$) and model ($\tilde{\lambda}^M(q_1)$) intensities of market orders as functions of the volume of the first limit $q_1$.}
\label{figure:MarketOrders-ModelIntensity-Level1}
\end{center}
\end{figure}

\begin{figure}
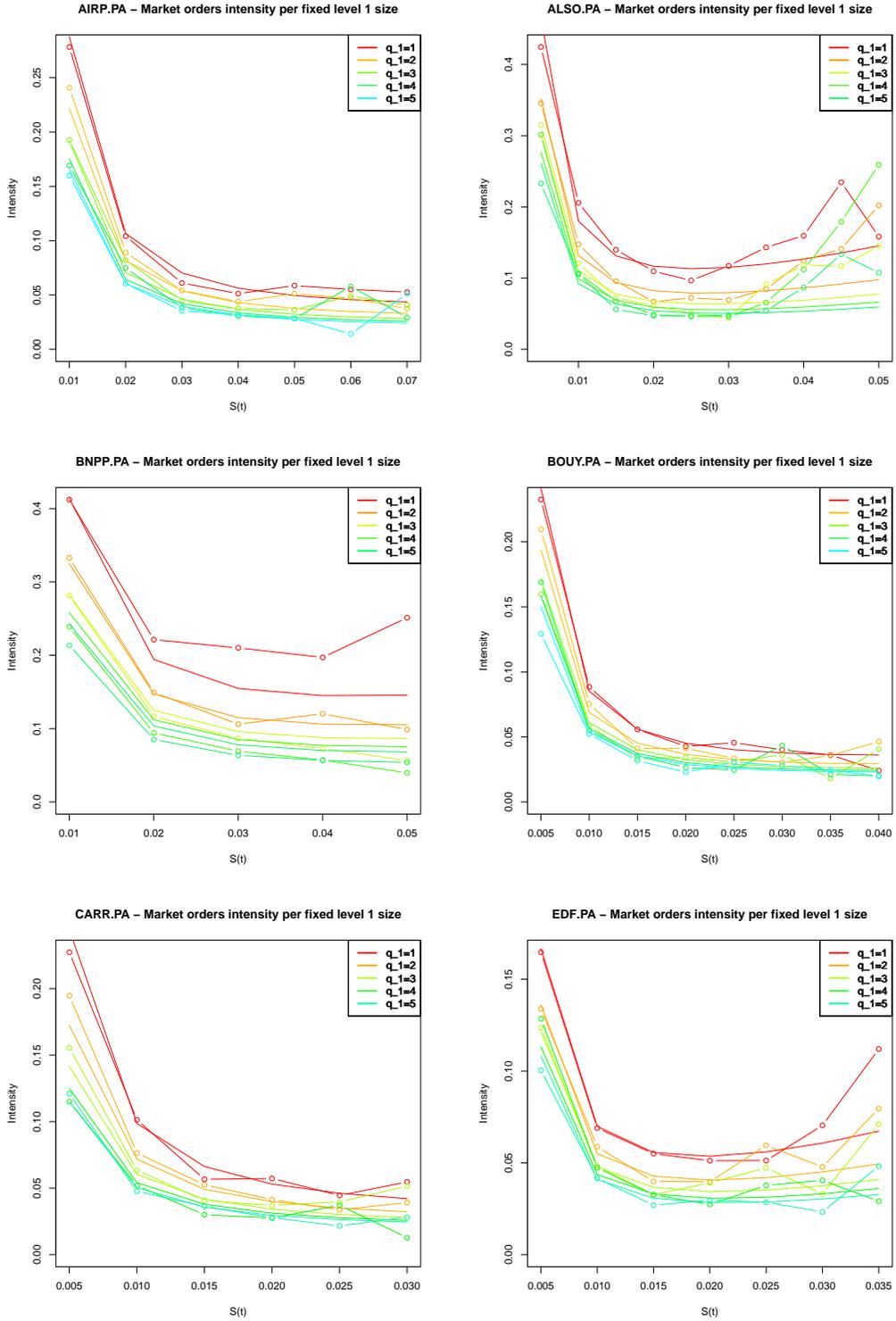

\begin{center}
\begin{tabular}{cc}
\includegraphics[width=0.4\textwidth, page=4]{{AIRP.PA-A-MarketOrders-SpreadDependence-0.99-20110117-20110128}.pdf}
&
\includegraphics[width=0.4\textwidth, page=4]{{ALSO.PA-A-MarketOrders-SpreadDependence-0.99-20110117-20110128}.pdf}
\\
\includegraphics[width=0.4\textwidth, page=4]{{BNPP.PA-A-MarketOrders-SpreadDependence-0.99-20110117-20110128}.pdf}
&
\includegraphics[width=0.4\textwidth, page=4]{{BOUY.PA-A-MarketOrders-SpreadDependence-0.99-20110117-20110128}.pdf}
\\
\includegraphics[width=0.4\textwidth, page=4]{{CARR.PA-A-MarketOrders-SpreadDependence-0.99-20110117-20110128}.pdf}
&
\includegraphics[width=0.4\textwidth, page=4]{{EDF.PA-A-MarketOrders-SpreadDependence-0.99-20110117-20110128}.pdf}
\end{tabular}
\caption{$q_1$-conditional intensities as functions of the spread. Lines with large dots represent empirical intensities and solid lines the fitted model intensities. Each $q_1$ level has one color.}
\label{figure:MarketOrders-ConditionalModelIntensity-Spread}
\end{center}
\end{figure}

\begin{figure}
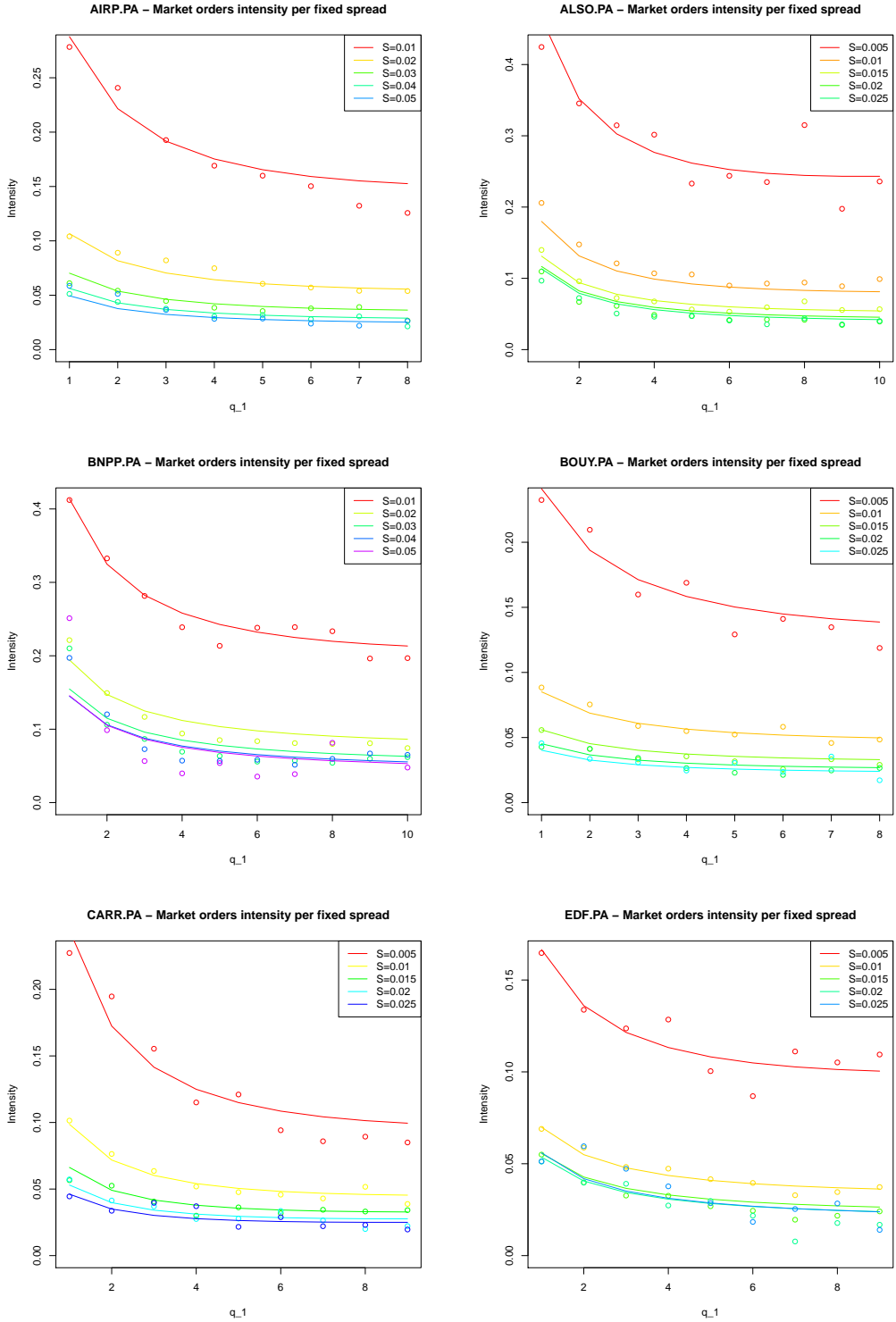

\begin{center}
\begin{tabular}{cc}
\includegraphics[width=0.4\textwidth, page=4]{{AIRP.PA-A-MarketOrders-Level1Dependence-0.9-20110117-20110128}.pdf}
&
\includegraphics[width=0.4\textwidth, page=4]{{ALSO.PA-A-MarketOrders-Level1Dependence-0.9-20110117-20110128}.pdf}
\\
\includegraphics[width=0.4\textwidth, page=4]{{BNPP.PA-A-MarketOrders-Level1Dependence-0.9-20110117-20110128}.pdf}
&
\includegraphics[width=0.4\textwidth, page=4]{{BOUY.PA-A-MarketOrders-Level1Dependence-0.9-20110117-20110128}.pdf}
\\
\includegraphics[width=0.4\textwidth, page=4]{{CARR.PA-A-MarketOrders-Level1Dependence-0.9-20110117-20110128}.pdf}
&
\includegraphics[width=0.4\textwidth, page=4]{{EDF.PA-A-MarketOrders-Level1Dependence-0.9-20110117-20110128}.pdf}
\end{tabular}
\caption{Spread-conditional intensities as functions of $q_1$. Dots represent empirical intensities and lines the fitted model intensities. Each spread level has one color.}
\label{figure:MarketOrders-ConditionalModelIntensity-Level1}
\end{center}
\end{figure}

We now provide several graphs to illustrate the fitting performance of the model.
We first plot for each stock the empirical intensity as a function of the spread ($\hat{\lambda}^M(S)$) and the "marginal" spread-dependent intensity $\tilde{\lambda}^M(S)$ computed by our model. This "marginal" represents the dependence on the spread when $q_1$ is distributed as in the sample, i.e. if it is computed with obvious notations as :
\begin{equation}
	\tilde{\lambda}^M(S) = \sum_{q} \lambda^M(t ; S, q) \mathbf P(q_1=q)
	\label{eq:LimitOrders-MarginalSpread}
\end{equation}
Similarly, we then plot for each stock the empirical intensity as a function of the level $q_1$ ($\hat{\lambda}^M(q_1)$) and the "marginal" $q_1$-dependent intensity $\tilde{\lambda}^M(q_1)$ computed by our model as :
\begin{equation}
	\tilde{\lambda}^M(q_1) = \sum_{s} \lambda^M(t ; s, q_1) \mathbf P(S=s)
	\label{eq:LimitOrders-MarginalLevel1}
\end{equation}
Results are given on Figure \ref{figure:MarketOrders-ModelIntensity-Spread} for the dependence on the spread and on Figure \ref{figure:MarketOrders-ModelIntensity-Level1} for the dependence on the volume at the best quote $q_1$.
The "marginal" intensities allow for a synthetic view of the modelling intensity.
In order to provide the reader with the full view of the fitting, we finally plot for each stock the spread-dependent empirical intensities given the volume $q_1$, and symmetrically the $q_1$-dependent intensities given the spead $S$. Results are plotted on Figures \ref{figure:MarketOrders-ConditionalModelIntensity-Spread} and \ref{figure:MarketOrders-ConditionalModelIntensity-Level1}, for each stock and each time for the first $5$ most probables occurrences of the variables.

\begin{figure}
\begin{center}
\begin{tabular}{cc}
\includegraphics[width=0.4\textwidth, page=1]{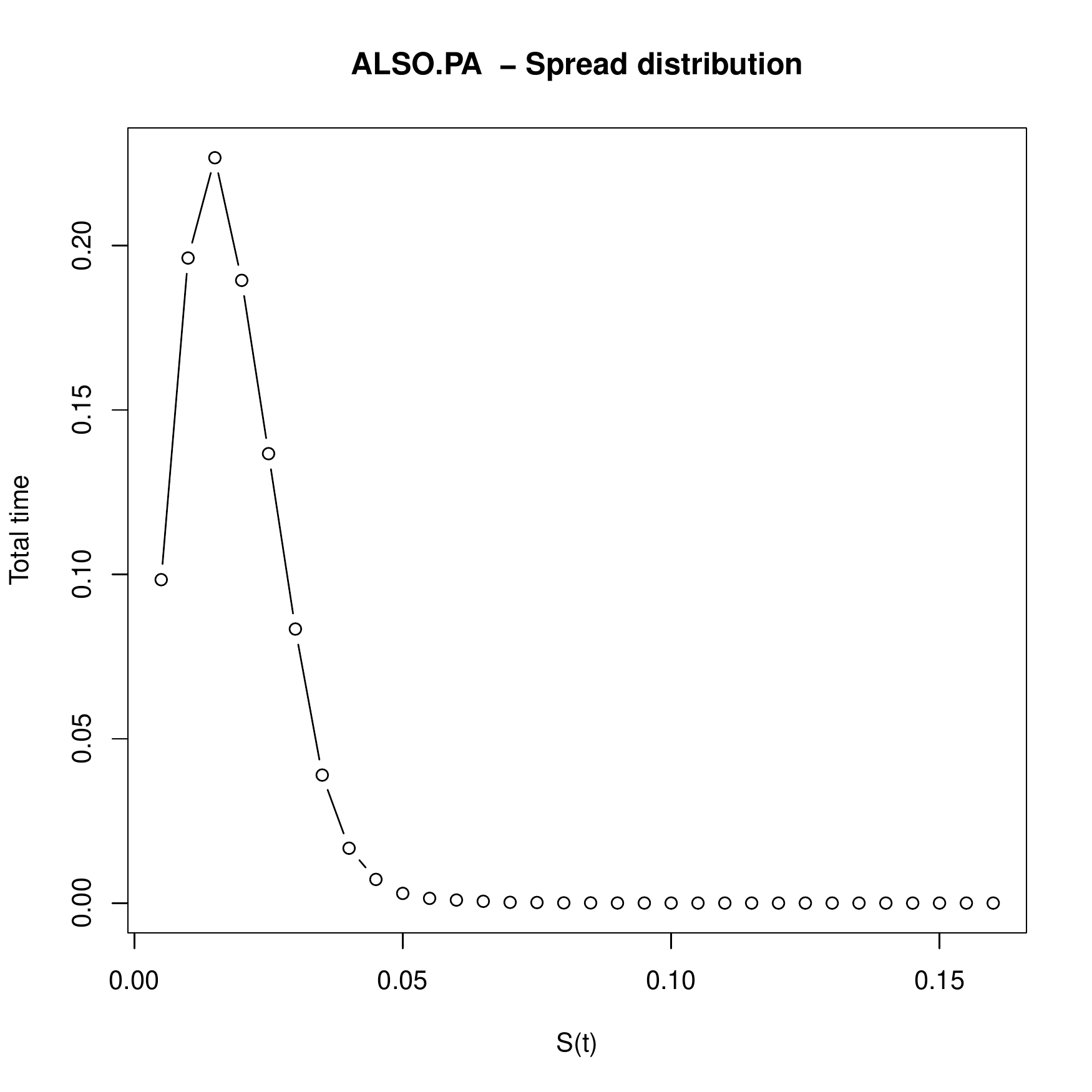}
&
\includegraphics[width=0.4\textwidth, page=1]{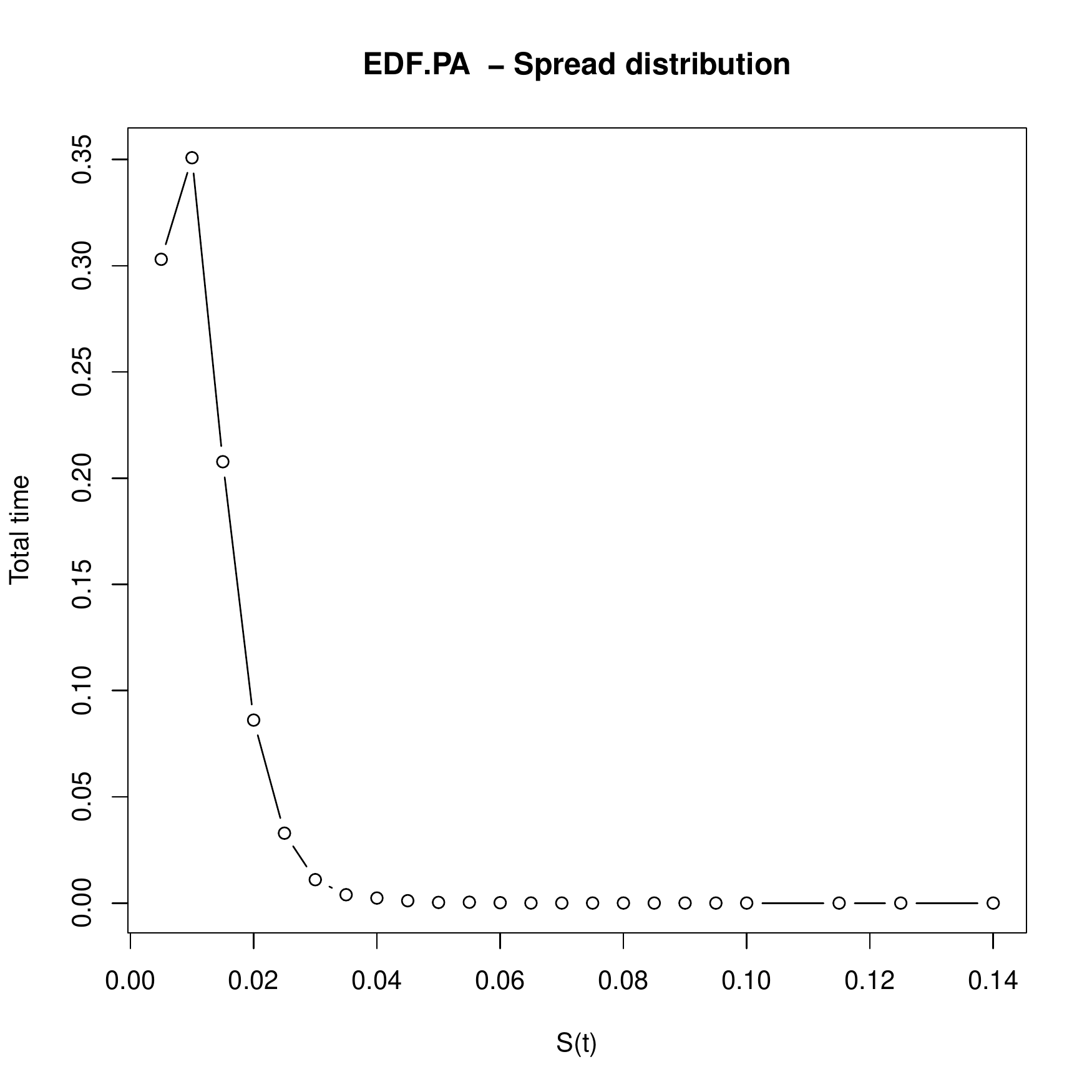}
\end{tabular}
\caption{Empirical distribution of the spread for two stocks, ALSO.PA (left) and EDF.PA(right).}
\label{figure:Spread-EmpiricalDistribution}
\end{center}
\end{figure}

Let us start with Figure \ref{figure:MarketOrders-ModelIntensity-Spread}. It turns out that the marginal fitting for the spread is always good, and even excellent for most stocks. It seems that it fails to catch the full extent of the increase of $\lambda^M$ observed the large values of the spread for some stocks (ALSO.PA, and to a lesser extent EDF.PA). It is however important to recall that high-spread values are very rare events. For two of the stocks under scrutiny here, we plot the empirical spread distribution in Figure \ref{figure:Spread-EmpiricalDistribution}.
This shows for example that for EDF.PA (Figure \ref{figure:MarketOrders-ModelIntensity-Spread}, bottom right), the last point on the right, which is the worst fit of the model, actually represents a few thousandths of the spread distribution.
It is therefore perfectly normal that the MLE estimation favors the main part of the distribution (left part of the graphs). This good fitting with respect to the spread is confirmed on Figure \ref{figure:MarketOrders-ConditionalModelIntensity-Level1} where each spread-conditional intensity is well modelled for each stock.

\begin{figure}
\begin{center}
\begin{tabular}{cc}
\includegraphics[width=0.4\textwidth, page=1]{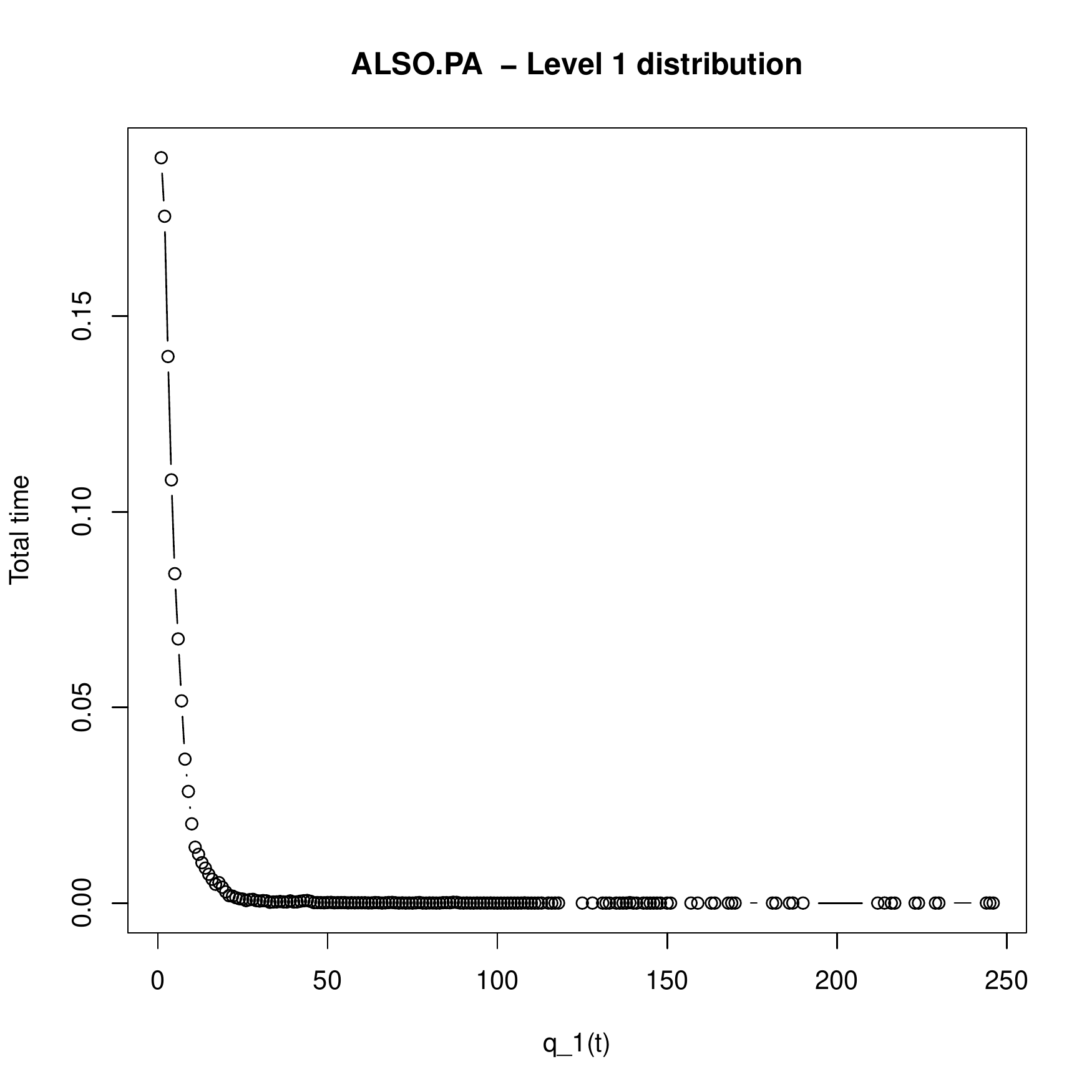}
&
\includegraphics[width=0.4\textwidth, page=1]{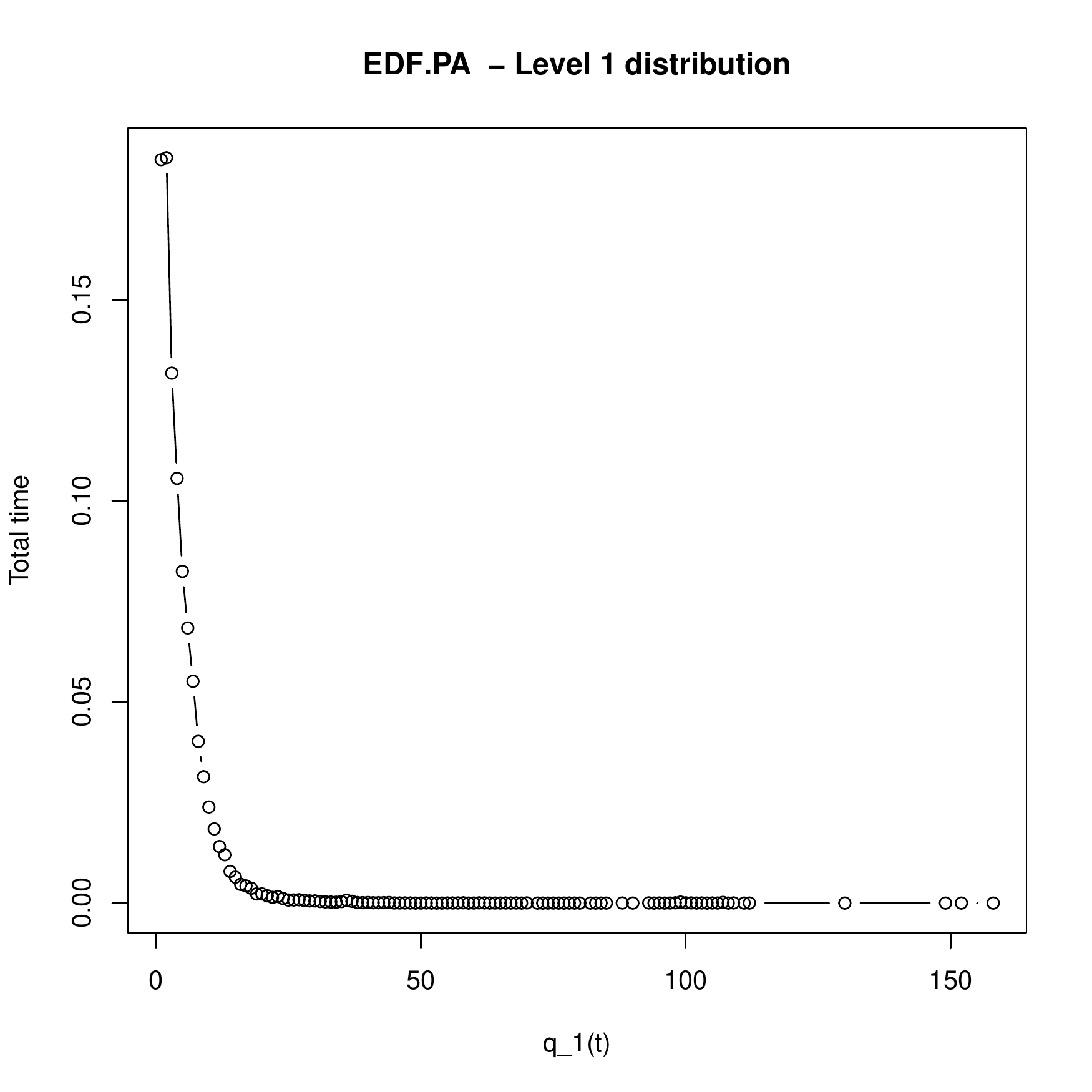}
\end{tabular}
\caption{Empirical distribution of $q_1$ for two stocks, ALSO.PA (left) and EDF.PA(right).}
\label{figure:Level1-EmpiricalDistribution}
\end{center}
\end{figure}

Continuing the analysis of the graphs, we observe on Figure \ref{figure:MarketOrders-ModelIntensity-Level1} that the quality of the fitting of the dependency on the volume $q_1$ seems a bit poorer. The model captures very well the decrease of the intensity as the volume $q_1$ increases, but the challenge here is that the empirical intensities are quite different from stock to stock : some decrease regularly, some faster at the beginning and then show a plateau.
This is also visible on Figure \ref{figure:MarketOrders-ConditionalModelIntensity-Spread} where larger levels of $q_1$ have less influence leading to the collapsing of the conditional intensities on the same curve.
The secondary role of larger values of $q_1$ is thus not surprising. There again, we show in Figure \ref{figure:Level1-EmpiricalDistribution} the empirical distribution of $q_1$ for two stocks for the sake of completeness.
The body of the distribution is clearly to the left, leaving less weight for the higher values. 

Therefore, the proposed model is overall a good fit, especially if we keep in mind that despite their differences we have managed to propose the same functional form for the dependence on the spread and the dependence on the volume at the best quote $q_1$.
We end this section by three modelling remarks, opening potential future works, and then move on to modelling of limit orders.
\begin{remark}
The form $\ln(1+q_1(t))$ is here preferred to $\ln(q_1(t))$ for flexibility as it allows for a normalized volume $q_1$ equal to zero. This is not the case in this paper since we have rounded above normalized volumes, so that $0$ is really $0$, not a small volume. But the difference being marginal, we keep the general (right-shifted) form.
\end{remark}
\begin{remark}
\label{remark:ModelUnification}
The likelihood analysis here is a conditional likelihood analysis given $S(t)$ and $q_1(t)$, or a regression analysis with these explanatory variables.
We discuss the modeling of limit orders and cancellations in the following sections, where a certain parametric model is introduced for each order.
Naturally, these models should be unified to describe the whole picture of all orders though we do not pursuit the integration of models in this paper.
\end{remark}
\begin{remark}
\label{remark:ModelSelection}
In the above construction of a model for the intensity, the exponential of a quadratic form of the logarithm of the variable is selected by the AIC criterion over an exponential of a quadratic form of the natural variable. Hence our choice that may not appear standard at first sight.
Furthermore, significance of every parameter suggests that we could introduce more explanatory variables and select a suitable model by a certain information criterion or a sparse estimation method. This is future work.
\end{remark}

\section{Limit orders}
\label{section:LimitOrders}

We now turn to the modelling of limit orders. Defining a limit order requires one dimension more than defining a market order :  its (limit) price has to be chosen upon submission.
We have decided to treat the two problems separately.
In a first subsection \ref{subsection:LimitOrders-Intensities}, we deal with the point process $N^L$ counting all limit orders (at any prices), with an instantaneous intensity $\lambda^L$.
The distribution of prices is assumed to be independently defined and will be discussed in the following subsection \ref{subsection:LimitOrders-Placement}.

\subsection{Modelling limit orders intensities}
\label{subsection:LimitOrders-Intensities}

Similarly to what we did for market orders, we choose two variables for our modelling. The price dimension is represented by the spread $S$. As for the "volume" dimension, we investigate the total volume available in the limit order book at the side of submission (more precisely the sum of all the liquidity available up to the tenth limit), denoted here $Q_{10}$. Since $\lambda^L$ deals with all limit orders, $Q_{10}$ appears obviously more relevant that $q_1$ as a modelling variable.

Following our modelling principles, we propose the following model for limit orders :
\begin{align}
	\lambda^L(t ; S(t), Q_{10}(t)) =  \exp \bigg[
	& \beta_0 + \beta_1 \ln(S(t)) + \beta_{11}[\ln(S(t))]^2 \nonumber
	+ \beta_2 \ln(1+Q_{10}) + \beta_{22}[\ln(1+Q_{10})]^2  \nonumber
	\\ & + \beta_{12}\ln(S(t))\ln(1+Q_{10}) \bigg].
\label{equation:LimitOrders-IntensityDefinition}
\end{align}
Here, we expect the intensity $\lambda^L$ to increase with the spread (by an argument exactly symmetric to the one we have used in Section \ref{section:MarketOrders}, see above).
We also expect it to increase with $Q_{10}$ decreases since by an expected stability mechanism, a global drop in the available volume should be an incentive to provide more liquidity.
As mentioned before, these monotonous variations guessed by "common financial sense" are only expected to be observed for frequent values of the modelling variables, since (rare) extreme values of the parameter are noisy and therefore difficult to characterize.

The model defined at Equation \eqref{equation:LimitOrders-IntensityDefinition} can be fitted by maximization of the likelihood.
It is straightforward to modify the formula given at Equation \eqref{eq:MarketOrders-NumericalLogLikelihood} to obtain the log-likelihood of the model, so we skip it for brevity.
The numerical results of the maximum likelihood  estimation are given in Table \ref{table:LimitOrders-fittedCoeffs}.
\begin{table}
\footnotesize
\begin{center}
\begin{tabular}{|l|rr|rr|rr|rr|rr|rr|}
\hline
ric & \multicolumn{1}{l}{$\beta_0$} & \multicolumn{1}{l|}{(std)} & \multicolumn{1}{l}{$\beta_1$} & \multicolumn{1}{l|}{(std)} & \multicolumn{1}{l}{$\beta_{11}$} & \multicolumn{1}{l|}{(std)} & \multicolumn{1}{l}{$\beta_2$} & \multicolumn{1}{l|}{(std)} & \multicolumn{1}{l}{$\beta_{22}$} & \multicolumn{1}{l|}{(std)} & \multicolumn{1}{l}{$\beta_{12}$} & \multicolumn{1}{l|}{(std)} \\ \hline
AIRP.PA & 5.772 & \textit{0.108} & -0.028 & \textit{0.031} & 0.048 & \textit{0.004} & -2.182 & \textit{0.041} & 0.263 & \textit{0.005} & 0.079 & \textit{0.005} \\ \hline
ALSO.PA & 14.516 & \textit{0.102} & 3.964 & \textit{0.030} & 0.443 & \textit{0.003} & -2.386 & \textit{0.035} & 0.248 & \textit{0.004} & -0.024 & \textit{0.005} \\ \hline
BNPP.PA & 6.472 & \textit{0.087} & 1.898 & \textit{0.040} & 0.285 & \textit{0.005} & -0.645 & \textit{0.018} & 0.101 & \textit{0.001} & 0.084 & \textit{0.004} \\ \hline
BOUY.PA & 17.042 & \textit{0.128} & 3.090 & \textit{0.041} & 0.296 & \textit{0.004} & -4.789 & \textit{0.040} & 0.507 & \textit{0.005} & -0.113 & \textit{0.007} \\ \hline
CARR.PA & 12.223 & \textit{0.150} & 0.699 & \textit{0.046} & 0.116 & \textit{0.005} & -4.635 & \textit{0.048} & 0.563 & \textit{0.005} & 0.079 & \textit{0.007} \\ \hline
EDF.PA & 15.176 & \textit{0.155} & 1.971 & \textit{0.047} & 0.164 & \textit{0.005} & -4.443 & \textit{0.052} & 0.456 & \textit{0.006} & -0.064 & \textit{0.007} \\ \hline
\end{tabular}
\caption{Fitted coefficients by maximum likelihood estimation for the intensity of limit orders.}
\end{center}
\label{table:LimitOrders-fittedCoeffs}
\end{table}
There again, standard deviations are provided to assess the quality of the fitting.

\begin{figure}
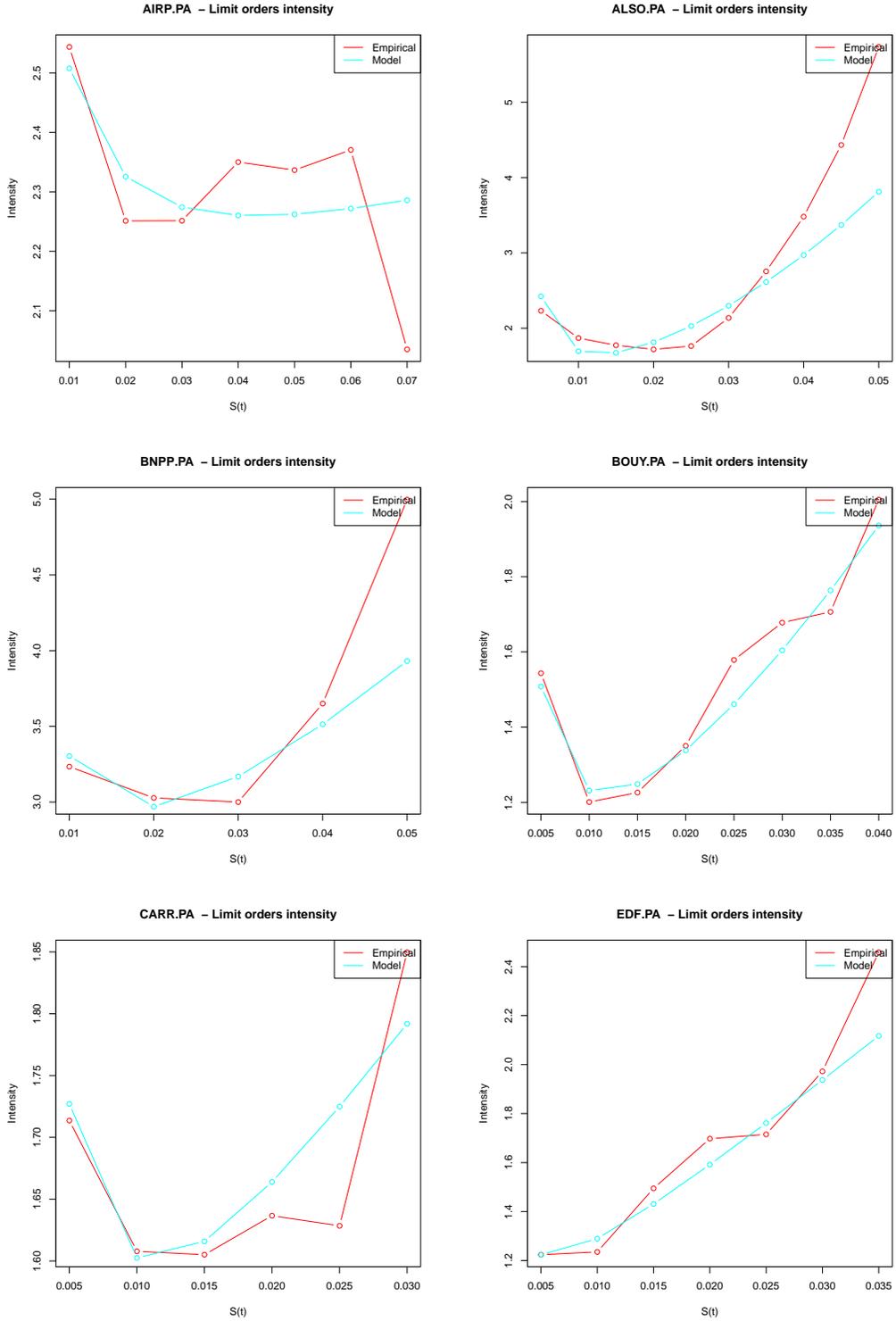

\begin{center}
\begin{tabular}{cc}
\includegraphics[width=0.4\textwidth, page=3]{{AIRP.PA-A-LimitOrders-SpreadDependence-0.99-20110117-20110128}.pdf}
&
\includegraphics[width=0.4\textwidth, page=3]{{ALSO.PA-A-LimitOrders-SpreadDependence-0.99-20110117-20110128}.pdf}
\\
\includegraphics[width=0.4\textwidth, page=3]{{BNPP.PA-A-LimitOrders-SpreadDependence-0.99-20110117-20110128}.pdf}
&
\includegraphics[width=0.4\textwidth, page=3]{{BOUY.PA-A-LimitOrders-SpreadDependence-0.99-20110117-20110128}.pdf}
\\
\includegraphics[width=0.4\textwidth, page=3]{{CARR.PA-A-LimitOrders-SpreadDependence-0.99-20110117-20110128}.pdf}
&
\includegraphics[width=0.4\textwidth, page=3]{{EDF.PA-A-LimitOrders-SpreadDependence-0.99-20110117-20110128}.pdf}
\end{tabular}
\caption{Empirical ($\hat{\lambda}^L(S)$) and model ($\tilde{\lambda}^L(S)$) intensities for limit orders as functions of the spread $S$.}
\label{figure:LimitOrders-ModelIntensity-Spread}
\end{center}
\end{figure}

\begin{figure}
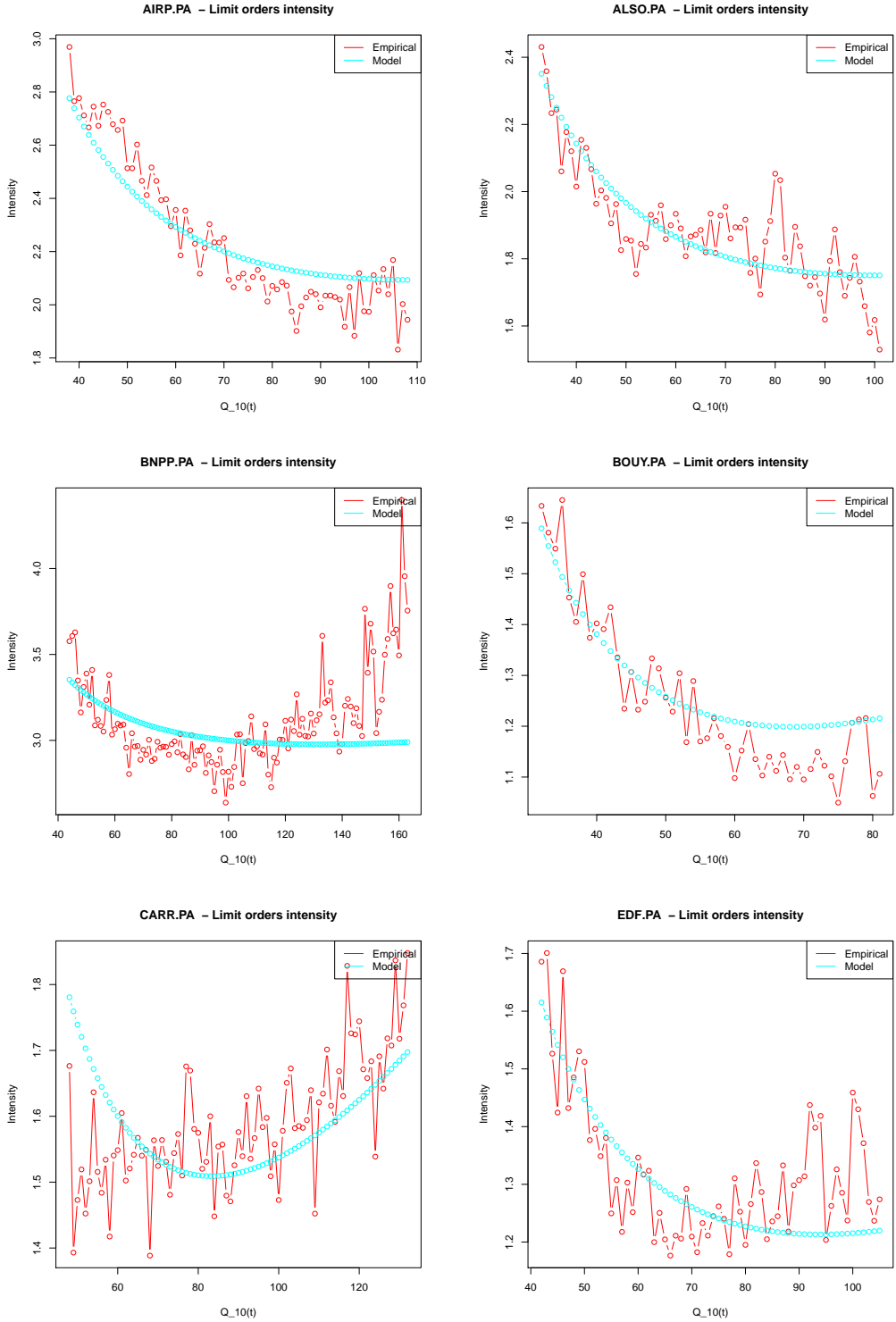

\begin{center}
\begin{tabular}{cc}
\includegraphics[width=0.4\textwidth, page=3]{{AIRP.PA-A-LimitOrders-tbvssDependence-0.8-20110117-20110128}.pdf}
&
\includegraphics[width=0.4\textwidth, page=3]{{ALSO.PA-A-LimitOrders-tbvssDependence-0.8-20110117-20110128}.pdf}
\\
\includegraphics[width=0.4\textwidth, page=3]{{BNPP.PA-A-LimitOrders-tbvssDependence-0.8-20110117-20110128}.pdf}
&
\includegraphics[width=0.4\textwidth, page=3]{{BOUY.PA-A-LimitOrders-tbvssDependence-0.8-20110117-20110128}.pdf}
\\
\includegraphics[width=0.4\textwidth, page=3]{{CARR.PA-A-LimitOrders-tbvssDependence-0.8-20110117-20110128}.pdf}
&
\includegraphics[width=0.4\textwidth, page=3]{{EDF.PA-A-LimitOrders-tbvssDependence-0.8-20110117-20110128}.pdf}
\end{tabular}
\caption{Empirical ($\hat{\lambda}^L(Q_{10})$) and model ($\tilde{\lambda}^L(Q_{10})$) intensities for limit orders as functions of the total book volume $Q_{10}$.}
\label{figure:LimitOrders-ModelIntensity-tbvss}
\end{center}
\end{figure}

We now provide graphical illustration of the quality of the fitting of the model.
One can straightforwardly adapt Equations \eqref{eq:LimitOrders-MarginalSpread} and \eqref{eq:LimitOrders-MarginalLevel1} to compute the "marginal" intensities of limit orders with respect to the spread and $Q_{10}$.
These are plotted on Figures \ref{figure:LimitOrders-ModelIntensity-Spread} and \ref{figure:LimitOrders-ModelIntensity-tbvss} where they are compared to the empirical intensities.
As for the dependence on the spread, we observe that the intensity $\hat{\lambda}^L(S)$ exhibits several shapes. There is indeed an increase for large spreads, as we expected, but for small spread we observe either a decrease or a plateau.
The model proposed is flexible enough to reproduce these shapes (except the unexpected drop for large spreads for AIRP.PA).
We could probably get better fits (for the eye) with some least-squares regression techniques, but the maximum-likelihood estimation chosen here emphasizes on the main body of the distribution, i.e. small spreads.

As for the dependence on the total volume available in the book on the side of submission $Q_{10}$, we observe that the intensity increases when the available liquidity decreases, as we expected. For some stocks (BNPP.PA or CARR.PA), we observe an increase when $Q_{10}$ increases above average, which the model is able to grasp.

Finally, the proposed model is once again a good fit, especially if we keep in mind that we have managed to propose the same functional form for the limit and market orders intensity, including both a price and a volume variable, following our modelling principle.

\subsection{Modelling the placement of limit orders}
\label{subsection:LimitOrders-Placement}

Modelling the placement of limit orders can a be difficult challenge. The support of any placement distribution is indeed state-dependent : in our model that distinguishes between three types of orders (limit, market, cancellation), one cannot submit a sell/buy limit order below/above the current best bid/ask. Such an order should be a market order.

With a simulation perspective, one can settle for a general distribution and then drop at the time of simulation any non-acceptable price (see Section \ref{section:Simulation}).
Using this technique, \citet{Mike2008} argued that the Student distribution centred around the current best quote is a good fit for the placement of limit orders (using data for the stock AstraZeneca on the London Stock Exchange).

In the same spirit, we will use continuous distributions on $\mathbb R$ to model the placement. $0$ will be the current best quote. We consider the placement distribution as a function on the \emph{continuous} variable price, and then integrate this density to obtain the discrete probability distribution of the placement of limit orders on the grid of integers numbers of ticksize.
If $\pi^L:\mathbb R\to\mathbb R_+$ is the continuous density of placement of limit orders and $\delta$ is the ticksize, then $\int_{(n-0.5)\delta}^{(n+0.5)\delta} \pi^L(u)\, du$ is the probability that the limit order is submitted a price $p=n\delta$.

We propose here two models. The first one is a generalized version of the \citet{Mike2008} proposition in which the limit orders are placed according the a location-scale version of the Student distribution:
\begin{equation}
	\pi^L(p ; \mu,\sigma, \nu) =  {\frac {\Gamma ({\frac {\nu +1}{2}})}{\Gamma ({\frac {\nu }{2}}){\sqrt {\pi \nu }}\sigma }}\left(1+{\frac {1}{\nu }}\left({\frac {x-\mu }{\sigma }}\right)^{2}\right)^{-{\frac {\nu +1}{2}}}
\label{equation:LimitOrders-Placement-StudentDefinition}
\end{equation}
This model interesting as it has only three parameters. However empirical data suggests that for some of the stocks we have studied placement of limit orders is often multi-modal. To our knowledge this observation has not been made before. One indeed observes a peak of submission at the best quote, and then another mode inside the book, a few ticks away from the best quote.
In order to reproduce this complex distribution we use a mixture of $G=3$ normal distributions:
\begin{equation}
	\pi^L(p ; G, \boldsymbol\mu, \boldsymbol\sigma, \boldsymbol\pi) =
	\sum_{i=1}^G \pi_i \phi(p ; \mu_i, \sigma_i),
\label{equation:LimitOrders-Placement-MixtureDefinition}
\end{equation}
where $\phi(\mu,\sigma)$ is the density of the Gaussian distribution with parameters $(\mu,\sigma)$.

The normal mixture model is fitted with the $\texttt{mclust}$ package of the $\texttt{R}$ language. The fitted parameters are given in Table \ref{table:LimitOrders-Placement-Mixture-fittedCoeffs}.
\begin{table}
\begin{center}
\begin{tabular}{|l|r|r|r|r|r|r|r|r|r|}
\hline
ric & \multicolumn{1}{l|}{$\pi_1$} & \multicolumn{1}{l|}{$\pi_2$} & \multicolumn{1}{l|}{$\pi_3$} & \multicolumn{1}{l|}{$\mu_1$} & \multicolumn{1}{l|}{$\mu_2$} & \multicolumn{1}{l|}{$\mu_3$} & \multicolumn{1}{l|}{$\sigma_1$} & \multicolumn{1}{l|}{$\sigma_2$} & \multicolumn{1}{l|}{$\sigma_3$} \\ \hline
AIRP.PA & 0.211 & 0.309 & 0.480 & 0.050 & 2.751 & 4.849 & 0.793 & 1.249 & 2.903 \\ \hline
ALSO.PA & 0.191 & 0.288 & 0.522 & 0.064 & 3.056 & 4.702 & 0.726 & 1.659 & 3.371 \\ \hline
BNPP.PA & 0.319 & 0.304 & 0.378 & 0.192 & 1.905 & 4.570 & 0.719 & 0.935 & 2.733 \\ \hline
BOUY.PA & 0.193 & 0.295 & 0.512 & 0.004 & 2.886 & 4.583 & 0.704 & 1.418 & 3.145 \\ \hline
CARR.PA & 0.235 & 0.283 & 0.483 & 0.159 & 2.405 & 4.560 & 0.768 & 1.214 & 2.695 \\ \hline
EDF.PA & 0.242 & 0.269 & 0.489 & 0.024 & 2.558 & 4.400 & 0.695 & 1.158 & 2.865 \\ \hline
\end{tabular}
\caption{Fitted parameters for the normal mixture model for the placement of limit orders. Parameters of the Gaussian distributions are expressed in number of ticksizes.}
\label{table:LimitOrders-Placement-Mixture-fittedCoeffs}
\end{center}
\end{table}
For all stocks, the fitted mixture model exhibits the same components. One Gaussian is centred on the best quote and very thin (standard deviation of two-third of a ticksize). This distribution accounts for roughly 20-25\% of the submitted limit orders, and helps modelling the peak of limit orders submitted at the best quote. Two other Gaussian distribution are further away in the book (roughly 2-3 and 4-5 ticks away from the best quote) help model the second mode observed and the more passive limit orders.

\begin{figure}
\begin{center}
\begin{tabular}{cc}
\includegraphics[width=0.4\textwidth, page=1]{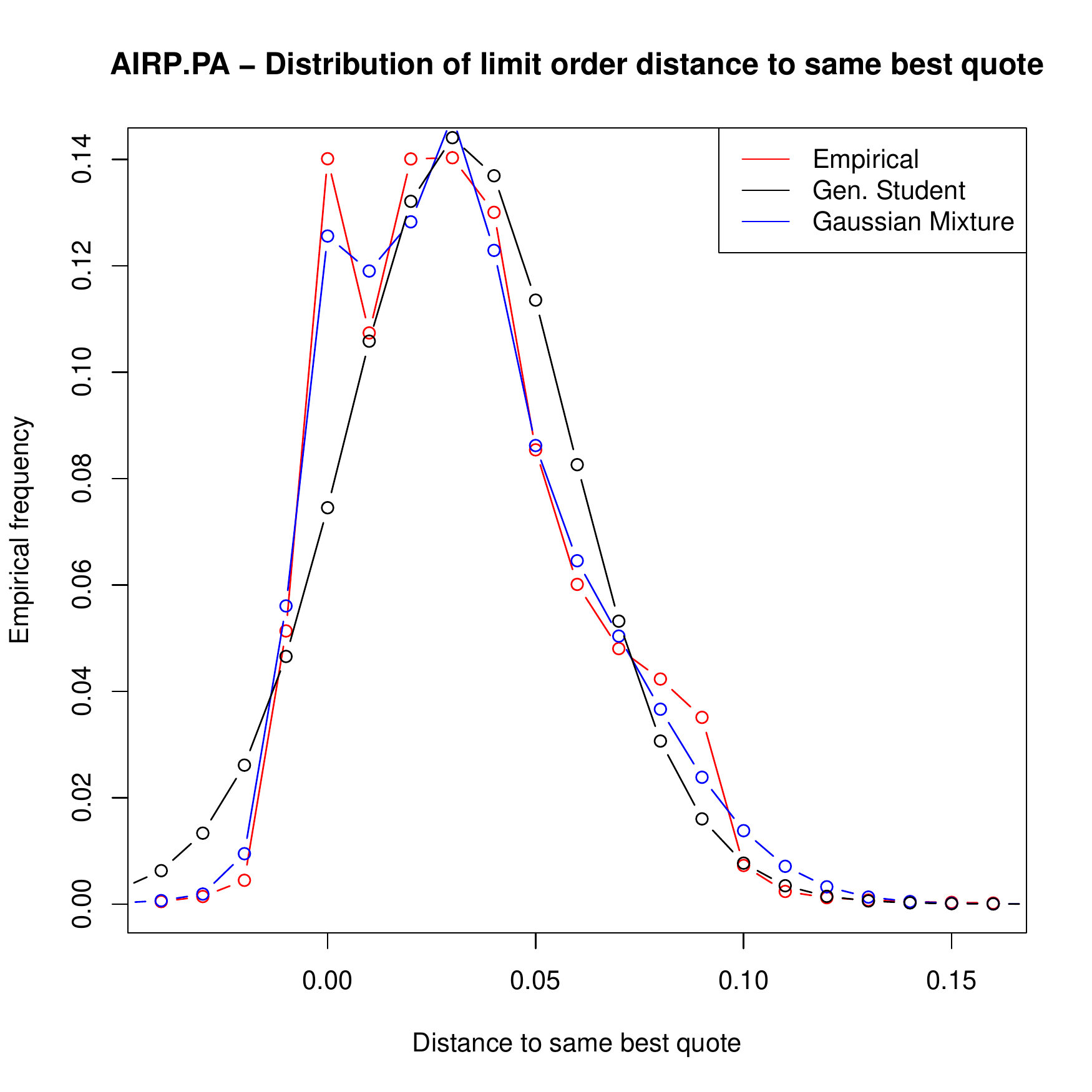}
&
\includegraphics[width=0.4\textwidth, page=1]{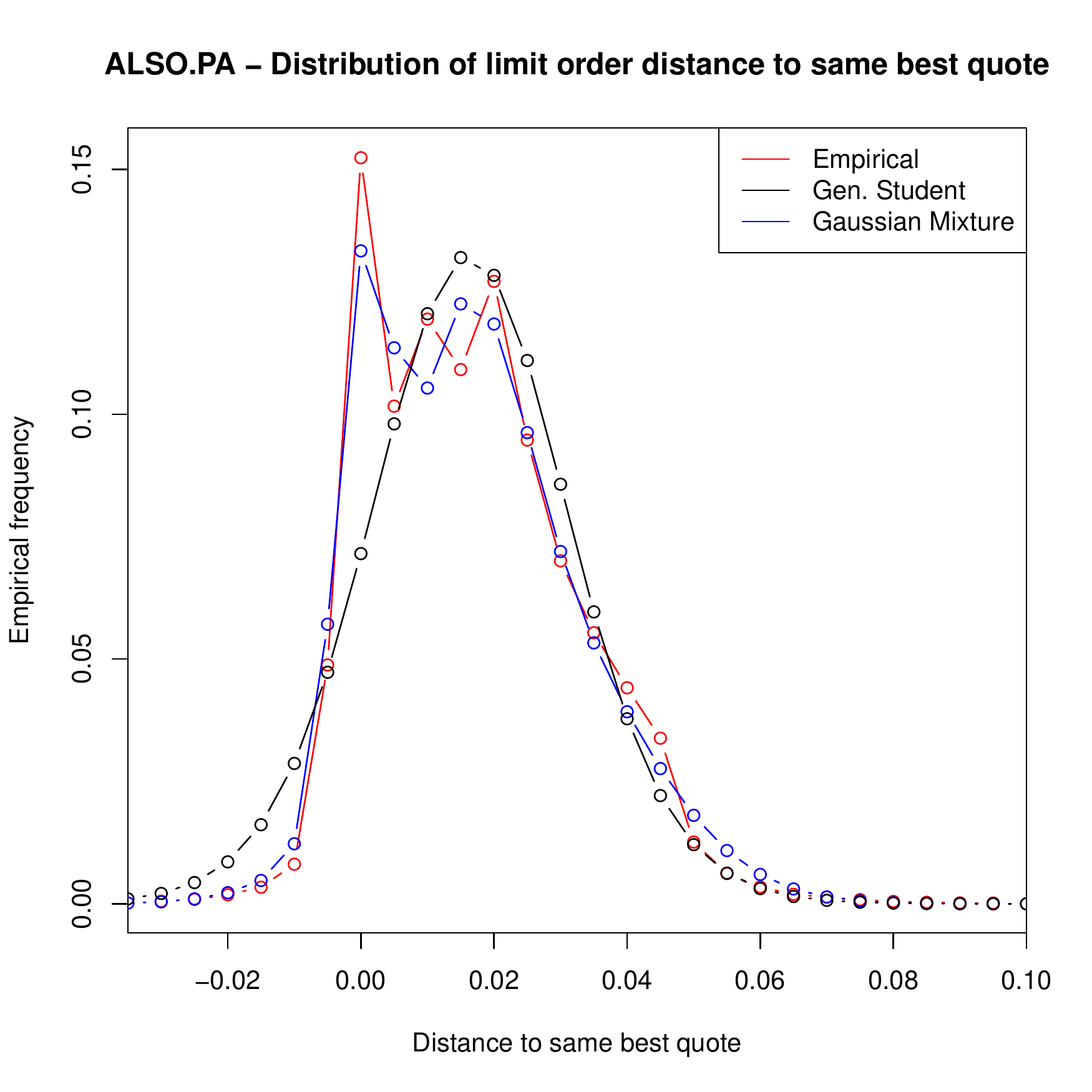}
\\
\includegraphics[width=0.4\textwidth, page=1]{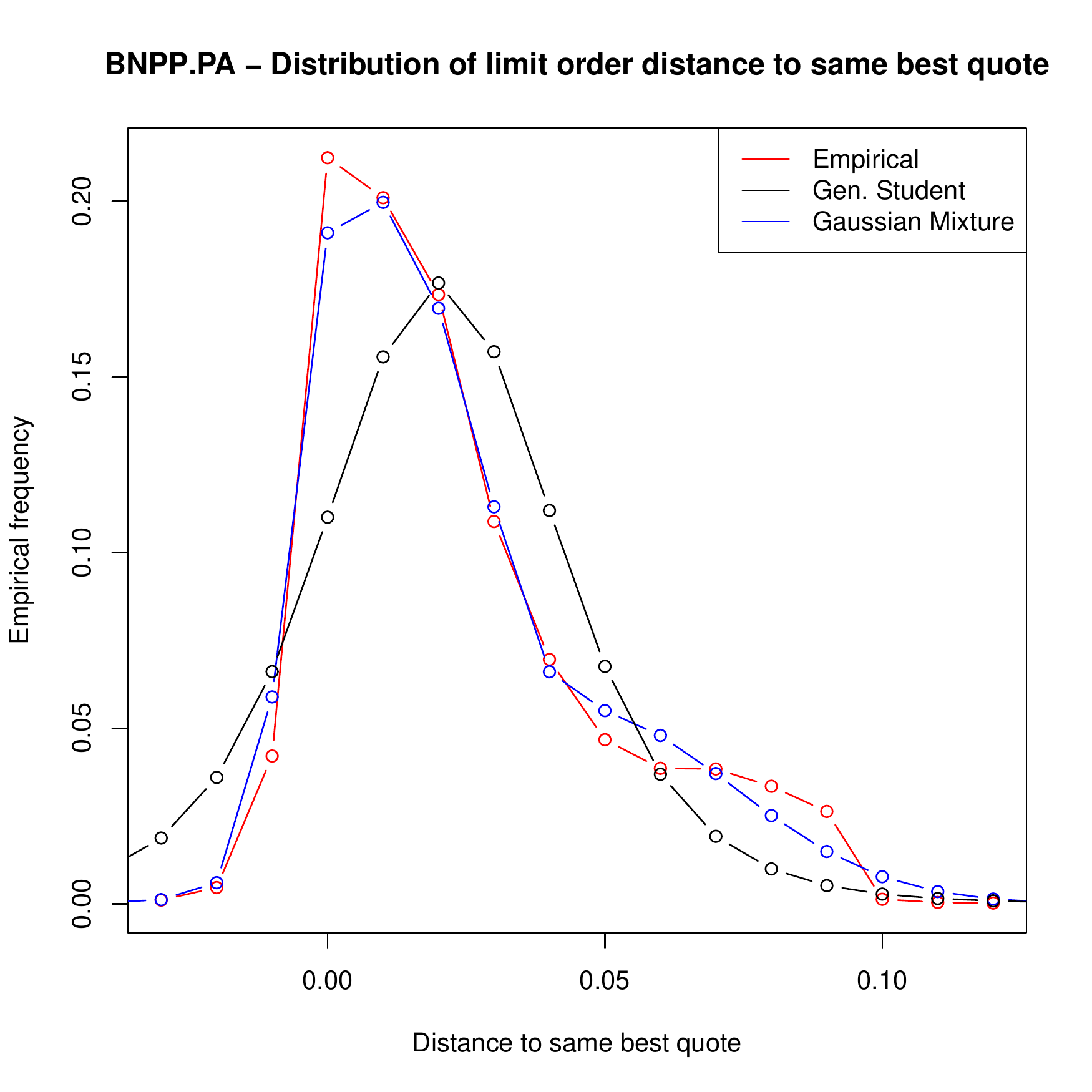}
&
\includegraphics[width=0.4\textwidth, page=1]{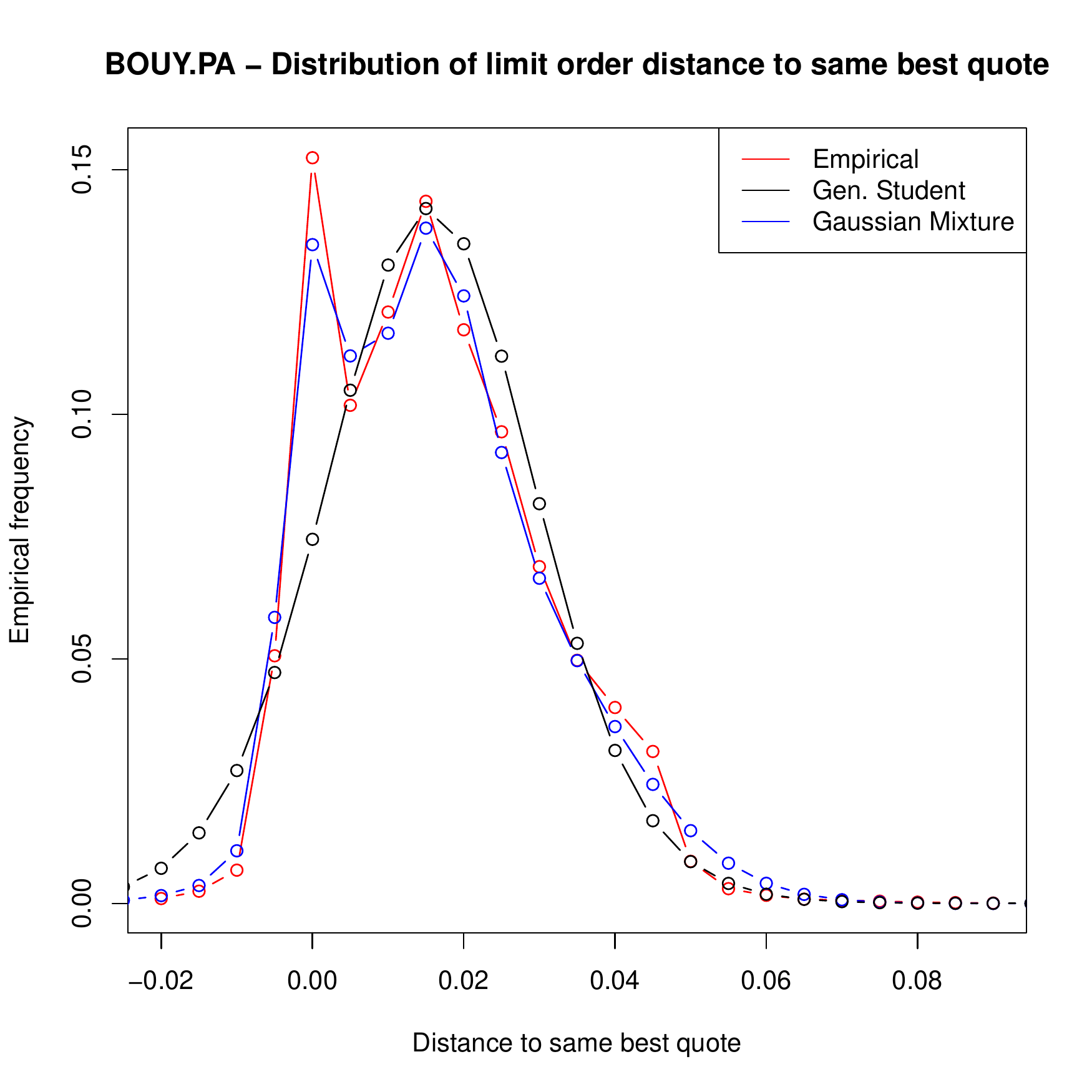}
\\
\includegraphics[width=0.4\textwidth, page=1]{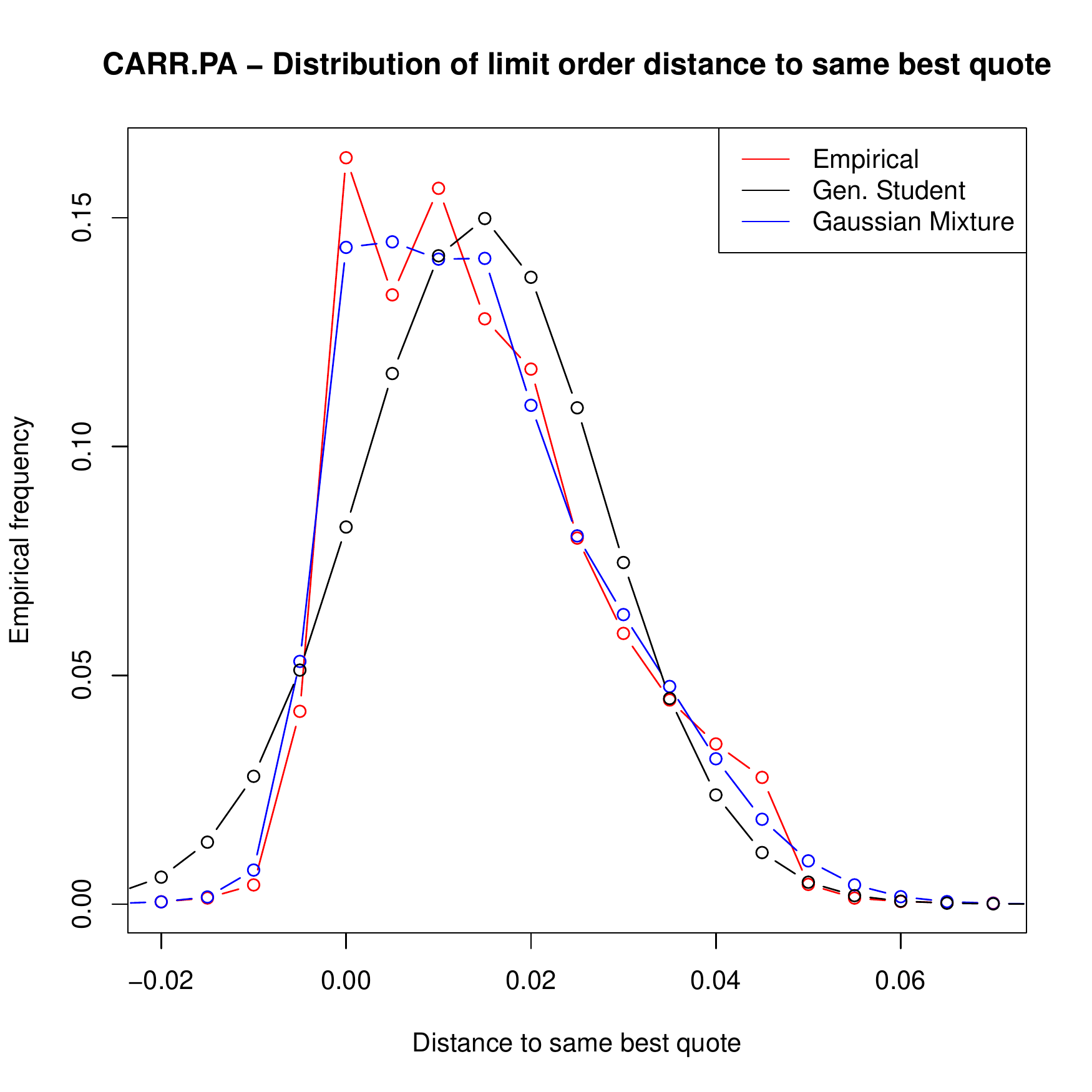}
&
\includegraphics[width=0.4\textwidth, page=1]{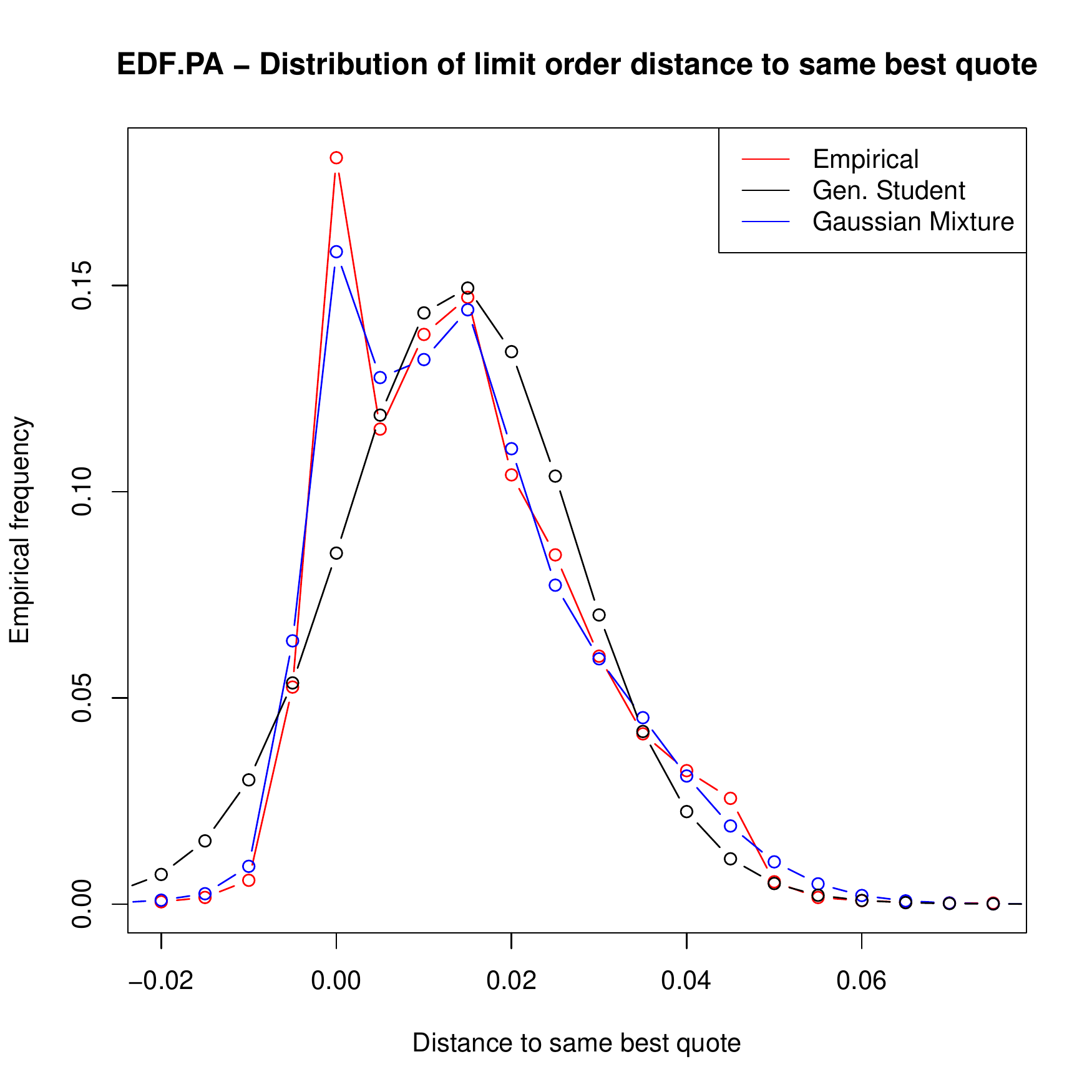}
\end{tabular}
\caption{Empirical and model distribution of the placement of limit orders.}
\label{figure:LimitOrders-Placement}
\end{center}
\end{figure}

In order to illustrate the quality of the fitting obtained, Figure \ref{figure:LimitOrders-Placement} plots the model distribution compared to the empirical one.
The fitted location-scale Student is given for comparison. This mono-modal distribution is in our sample centred on the maximum inside the book, a few ticks away from the best. As a result, it underestimates on the one hand the number of orders submitted at the best quote, but on the on the other hand it overestimates the number of aggressive orders submitted inside the spread.

\begin{remark}
\label{remark:MultimodalitySpread}
We observe that the multi-modality of the placement of limit orders strongly depends on the observed spread. 
It is usually stronger for small spreads, and disappears for larger spread. 
This can be interpreted as follows.
When the spread is smaller than usual, the market participants anticipate its widening, thus providing liquidity a few ticks inside the book besides the usual liquidity provided at the best quote. Hence the appearance of two peaks in the distribution on the placement of limit orders, and the strong multi-modality.
When the spread is large, market participants anticipate its tightening, thus providing more liquidity close to the best quote, hence the disappearance of the multi-modality.

It is easy to generalize our model given at Equation \eqref{equation:LimitOrders-Placement-MixtureDefinition} to a spread-dependent model, by splitting our sample according to the observed spread and then fitting spread-dependent parameters:
\begin{equation}
	\pi^L(p ; G, \boldsymbol\mu, \boldsymbol\sigma, \boldsymbol\pi) =
	\sum_{i=1}^G \pi_i(S) \phi(p ; \mu_i(S), \sigma_i(S)).
\label{equation:LimitOrders-Placement-MixtureDefinition-SpreadDependent}
\end{equation}
This would increase the number of parameters of the model but allow for a better flexibility in the modelling of the placement of limit orders.
With the simulation of Section \ref{section:Simulation} in mind and given the good performances of the proposed fit, we stick, at least for now, to the unconditional model.
\end{remark}

\section{Cancellations of pending orders}
\label{section:Cancellations}

Cancellations are different from the two previous types of orders studied (limit and market) because they are not a message to buy or sell some shares on the market, but a message to cancel a previous message to buy or sell some shares. For example, we cannot model the placement of cancellations as we did for the limit orders, since we can only cancel orders at prices where some orders at actually standing in the book.
We thus adopt a completely different type of modelling for cancellations.

The first choice of modelling is that we do not model the intensity of submission of cancellation, but we model instead the lifetime of pending limit orders.
One reason for this choice is that cancellations ensures the stability of the system. Cancellation process is intimately linked to the limit submission process. By defining an autonomous state-dependent cancellation process, we introduce a risk of instability in the model.
The choice of the lifetime of orders as the main variable is thus a safe choice.
Its drawback however is that it is a very difficult parameter to estimate.
Our trades and quotes database does not provide a unique identifier for each order, thus when we observe a cancellation we do no know for sure which limit order has been cancelled. We can narrow it down by selecting only limit orders with the volume and price equal to the one cancelled, but this identification does not necessarily return a unique match.
Finally, even if we perform the above algorithm with some selection rules, the obtained distribution is not necessarily easy to characterize.
As an example, on the stock AIRP.PA on January 17th, 2011, the above algorithm gives an empirical distribution of lifetimes with median of 5.2 seconds, and a mean of 89.7 seconds.

We choose to compute the average lifetime of an order so that a basic order book model with Poisson intensities would have an average total liquidity in the book equal to the empirical observation.
More precisely, \cite{MuniToke2015} shows that in an order book with Poisson arrival of market orders with intensity $\lambda^M\in\mathbb R_ +$ and average size $\sigma^M$, Poisson arrival of limit orders with intensity $\lambda^L\in\mathbb R_ +$ and average size $\sigma^L$, and a lifetime of pending limit orders exponentially distributed with parameter $\theta^{-1}$, the expected total liquidity $Q$ available in the book is
\begin{equation}
    Q = \sigma^M \left(\frac{\nu}{q} - \delta 
    + \frac{\delta q^{\frac{\nu}{1-q}}}{{}_2F_1\left(\delta,-\frac{\nu}{1-q},1+\delta,1-q\right)} \right)
\label{equation:Cancellations-AverageShape}
\end{equation}
where $\nu = \frac{\lambda^L}{\theta}$, $\delta=\frac{\lambda^M}{\theta}$, $q=\frac{\sigma^M}{\sigma^L}$ and ${}_2F_1$ is the hypergeometric function.
It is easy to numerically optimize $\theta$ so that $Q$ given in the equation above is equal to its empirical counterpart.

The second choice of modelling deals with the "placement" of the cancellations. 
\cite{Mike2008} has proposed a three-variable model to determine the placement of cancellations, based on the distance to the best quote, the total liquidity available and the imbalance.
We here propose a new model efficient one-parameter model to choose which pending order is to be cancelled.
We introduce as modelling variable the "priority index".
We firstly define the "priority volume" of a limit order as the sum of all the sized of pending limit orders standing ahead in the queue, i.e. at a better price or at the same price but with time priority. If a limit order is the oldest order standing at the best quote, then it will be executed first when a market order arrives, its priority volume is thus zero.
One may expect that the probability to be cancelled decreases with the priority volume, but that would be ignoring the fact that most of the activity occurs around the best quotes.

\begin{figure}
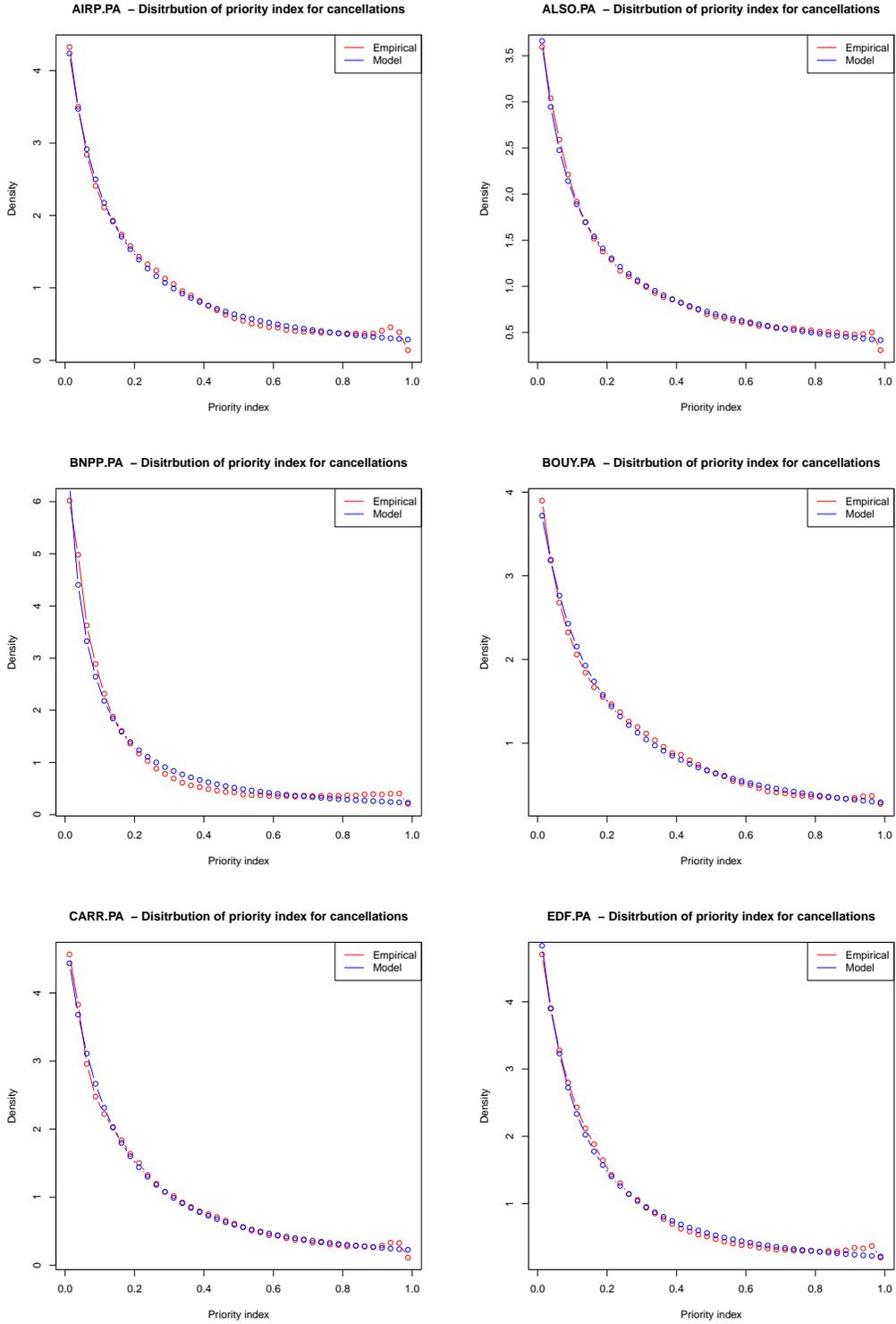

\begin{center}
\begin{tabular}{cc}
\includegraphics[width=0.4\textwidth, page=2]{{AIRP.PA-A-CancelOrders-Placement-Tpv-20110117-20110128}.pdf}
&
\includegraphics[width=0.4\textwidth, page=2]{{ALSO.PA-A-CancelOrders-Placement-Tpv-20110117-20110128}.pdf}
\\
\includegraphics[width=0.4\textwidth, page=2]{{BNPP.PA-A-CancelOrders-Placement-Tpv-20110117-20110128}.pdf}
&
\includegraphics[width=0.4\textwidth, page=2]{{BOUY.PA-A-CancelOrders-Placement-Tpv-20110117-20110128}.pdf}
\\
\includegraphics[width=0.4\textwidth, page=2]{{CARR.PA-A-CancelOrders-Placement-Tpv-20110117-20110128}.pdf}
&
\includegraphics[width=0.4\textwidth, page=2]{{EDF.PA-A-CancelOrders-Placement-Tpv-20110117-20110128}.pdf}
\end{tabular}
\caption{Empirical and model distribution of the placement of cancellations as a function of the priority index.}
\label{figure:Cancellations-Placement}
\end{center}
\end{figure}

Let us now define the "priority index" $\xi$ of a pending limit order as the ratio of the "priority volume" defined above over the total volume available in the book (on the same side). Obviously $\xi\in[0,1]$. $\xi$ can be used as a indicator of placement of cancellations inside the book.
As for the empirical estimation of $\xi$ however, our data does not allow for the unique tracking of individual orders. We know the price of an order, but not exactly where the order is inside the sub-queue of all orders at this price (at least not without further algorithmic development). We thus compute the priority volume as the total liquidity available at better prices plus half the liquidity available at the same price, i.e. we act as if the cancelled order were in the middle of the queue. This allows for an easy estimation of $\xi$ on our data.
It turns out that the distribution of cancellations as a function of $\xi$ is remarkably smooth.
Some empirical results are given below. 
We propose to model it with a scaled truncated power law distribution, i.e. we have the following model for the density of the cancellation "placement" $\pi^C:[0,1]\to\mathbb R_ +$:
\begin{equation}
	\pi^C(\xi) = \frac{\sigma(\alpha+1)}{(1+\sigma)^{\alpha+1}-1} (1+\sigma \xi)^{\alpha}.
\label{equation:Cancellations-TpvDensity}
\end{equation}
The log-likelihood of a sample $(\xi_1,\ldots,\xi_N)$ is straightforwardly computed as
\begin{equation}
	\mathcal{L}(\alpha, \sigma) = N \log\left( \frac{\sigma(\alpha+1)}{(1+\sigma)^{\alpha+1}-1} \right)
	+ \alpha \sum_{i=1}^N \log(1+\sigma \xi_i),
\end{equation}
which can be numerically maximized using the \texttt{mle2} routine of the \texttt{bbmle} package. 
Numerical results of the maximum-likelihood estimation are given in Table \ref{table:Cancellations-Placement-fittedCoeffs}.
\begin{table}[htbp]
\begin{center}
\begin{tabular}{|l|r|r|r|r|}
\hline
ric & \multicolumn{1}{l|}{$\alpha$} & \multicolumn{1}{l|}{(std)} & \multicolumn{1}{l|}{$\sigma$} & \multicolumn{1}{l|}{(std)} \\ \hline
AIRP.PA & -1.378 & \textit{0.008} & 6.760 & \textit{0.095} \\ \hline
ALSO.PA & -0.876 & \textit{0.005} & 13.101 & \textit{0.241} \\ \hline
BNPP.PA & -1.256 & \textit{0.004} & 16.014 & \textit{0.132} \\ \hline
BOUY.PA & -1.561 & \textit{0.017} & 4.412 & \textit{0.098} \\ \hline
CARR.PA & -1.775 & \textit{0.015} & 4.684 & \textit{0.078} \\ \hline
EDF.PA & -1.694 & \textit{0.013} & 5.749 & \textit{0.090} \\ \hline
\end{tabular}
\end{center}
\caption{Parameters for the placement of cancellations obtained by numerical minimization of the loglikelihood.}
\label{table:Cancellations-Placement-fittedCoeffs}
\end{table}
Illustrations of the quality of the fit are provided on Figure \ref{figure:Cancellations-Placement}.
Table and figures all show an excellent agreement between the model and the empirical data for all the stocks studied.

\section{A market simulator with state-dependent order flows}
\label{section:Simulation}

We show the benefits of our model by fitting it to daily empirical data and simulating it. 
Simulating a "realistic" limit order book is a quite complex task given the many parameters involved and the somewhat complex time-priority execution mechanism to be implemented.
Several results have previously been obtained, for example in \cite{Gatheral2010,MuniToke2011}. Some key elements for basic simulation can be found in \cite{Abergel2015}.

\subsection{Market simulator}
We build a market simulator with four agents. Two "liquidity providers" submit (and cancel) limit orders, one on the ask side and another on the bid side. Two "liquidity takers" submit market orders, one on the ask side and another on the bid side.
We choose to simulate here a symmetric limit order book, i.e. both providers share the same parameters, and both takers share the same parameters.

Liquidity providers submit limit orders with the intensity $\lambda^L(S, Q_{10})$ defined in Equation \eqref{equation:LimitOrders-IntensityDefinition}.
The distribution of the sizes of the limit orders is exponentially distributed with parameters $\frac{1}{\hat{\sigma}^L}$ where $\hat{\sigma}^L$ is the median of the empirical sizes of limit orders.
The distribution of the prices of the limit orders is defined by our Gaussian mixture model given by Equation \eqref{equation:LimitOrders-Placement-MixtureDefinition}.

Liquidity takers submit market orders with the intensity $\lambda^M(S, q_1)$ defined in Equation \eqref{equation:MarketOrders-IntensityDefinition}. The distribution of the sizes of the limit orders is exponentially distributed with parameters $\frac{1}{\hat{\sigma}^M}$ where $\hat{\sigma}^M$ is the median of the empirical sizes of limit orders.

Finally, cancellations in the order book occur with an intensity proportional to the available liquidity, i.e. $\lambda^C= Q\theta$ where $Q$ is the total number of orders and $\theta$ is determined by the procedure detailed in Section \ref{section:Cancellations} and Equation \eqref{equation:Cancellations-AverageShape}. When a cancellation occurs, a random priority index $\bar{\xi}$ is drawn according to the distribution with density $\pi^C$ given at Equation \eqref{equation:Cancellations-TpvDensity}.
This distribution is easy to simulate given its inverse cumulative distribution function $(\Pi^C)^{-1}$:
\begin{equation}
	(\Pi^C)^{-1}(x) = \frac{1}{\sigma} \left[ \left[ \left((1+\sigma)^{\alpha+1}-1\right)x+1\right]^{\frac{1}{\alpha+1}}-1\right].
\end{equation}
The order cancelled is then the first one that has a priority index greater or equal to $\bar{\xi}$.

\subsection{Poisson simulator reference}
To provide a reference simulation, we simulate a standard Poisson model. This reference model has the same agents, the same distributions of sizes of limit and market orders, and the same cancellation intensity proportional to the liquidity available.
However, all agents submit their orders according to a homogeneous Poisson process with a constant intensity fitted by MLE estimation. The placement of limit orders is done according to the location-scale Student distribution given in Equation \eqref{equation:LimitOrders-Placement-StudentDefinition}.
Finally, the cancellation is purely zero-intelligence in the sense that the chosen order when a cancellation occurs is uniformly selected in the book.

\subsection{Simulation results}

We fit our model for each stock of our sample, and using one day of trading.
Since we simulate a symmetric limit order book, we aggregate bid and ask order flows in one sample for the fitting.
We have made the full simulation of our model for each of the first two days of the sample, but for the sake of brevity, we show in this section the results for only one day, January 18th, 2011. Results for the other day tested are exactly similar. 
The sample used for fitting is smaller than the full one (ten days) used in the previous sections to derive the functional shapes of the intensities and distributions of our model. This may lead to potentially noisier estimates of our model, but for practical purposes one trading day is a convenient unit of time, hence this choice.

The simulator (and the reference Poisson simulator) is then run to produce exactly one day of trading data (i.e. the same length as the fitting sample). We then analyse the simulated data and compare it to the empirical observations.

\begin{figure}
\begin{center}
\begin{tabular}{cc}
\includegraphics[width=0.4\textwidth, page=1]{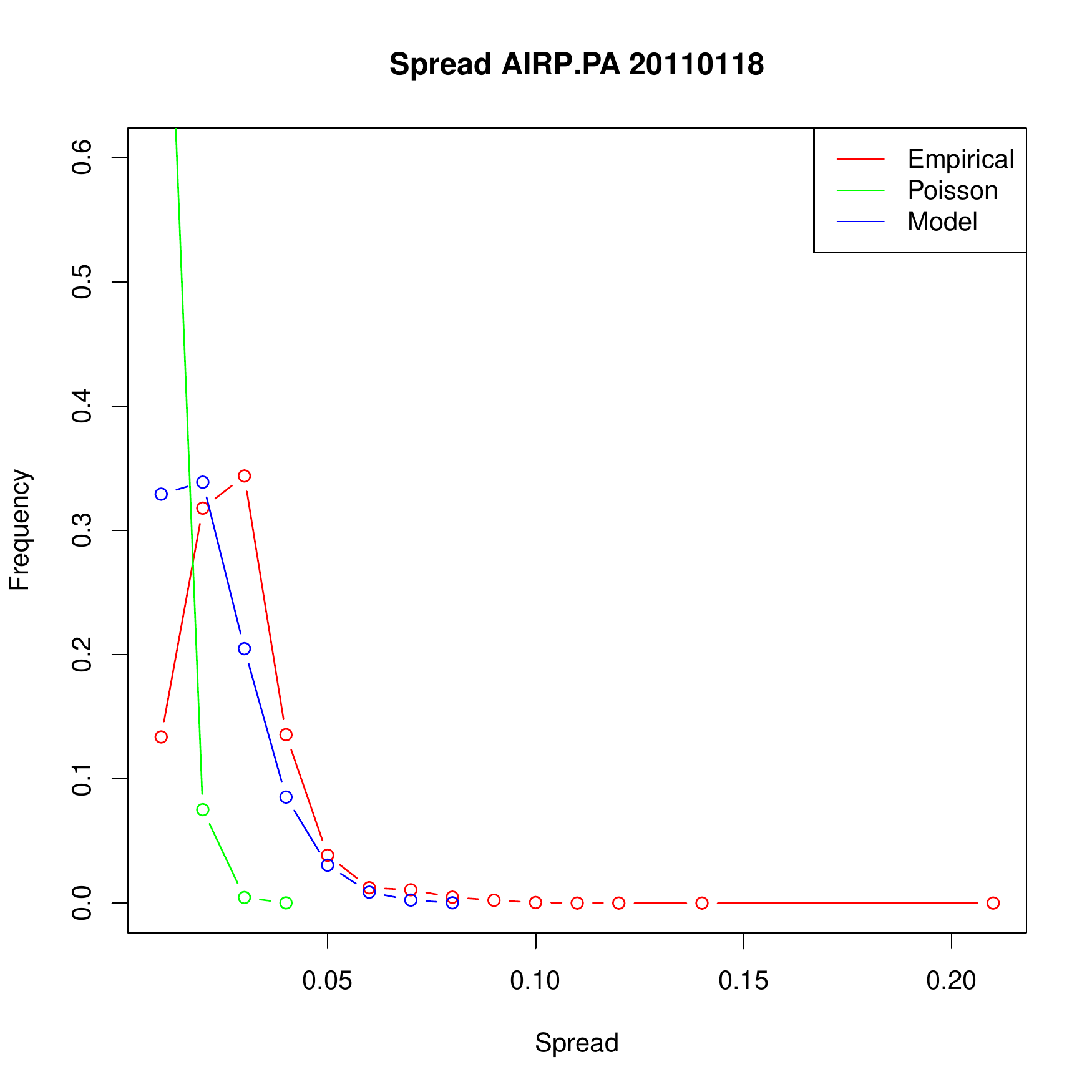}
&
\includegraphics[width=0.4\textwidth, page=1]{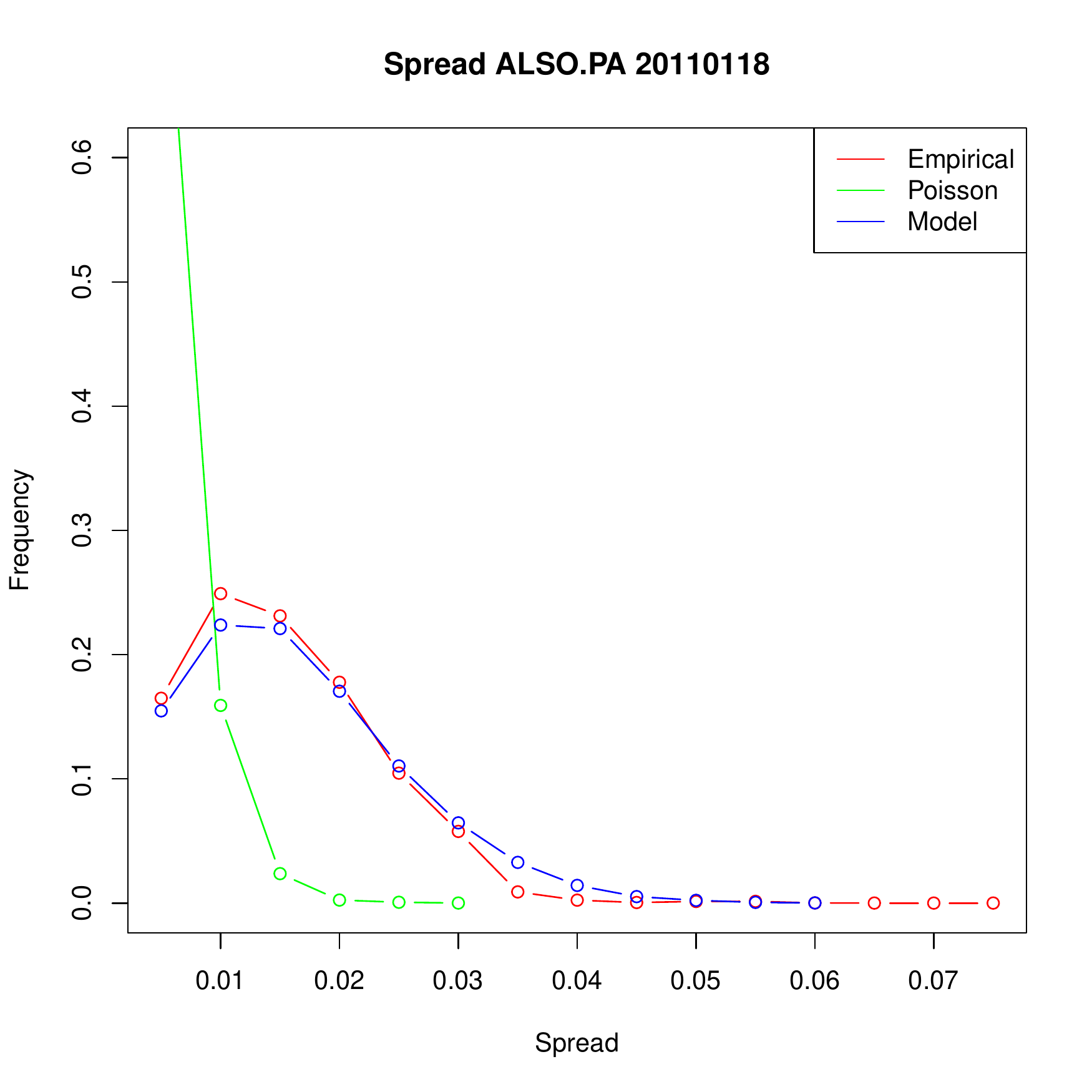}
\\
\includegraphics[width=0.4\textwidth, page=1]{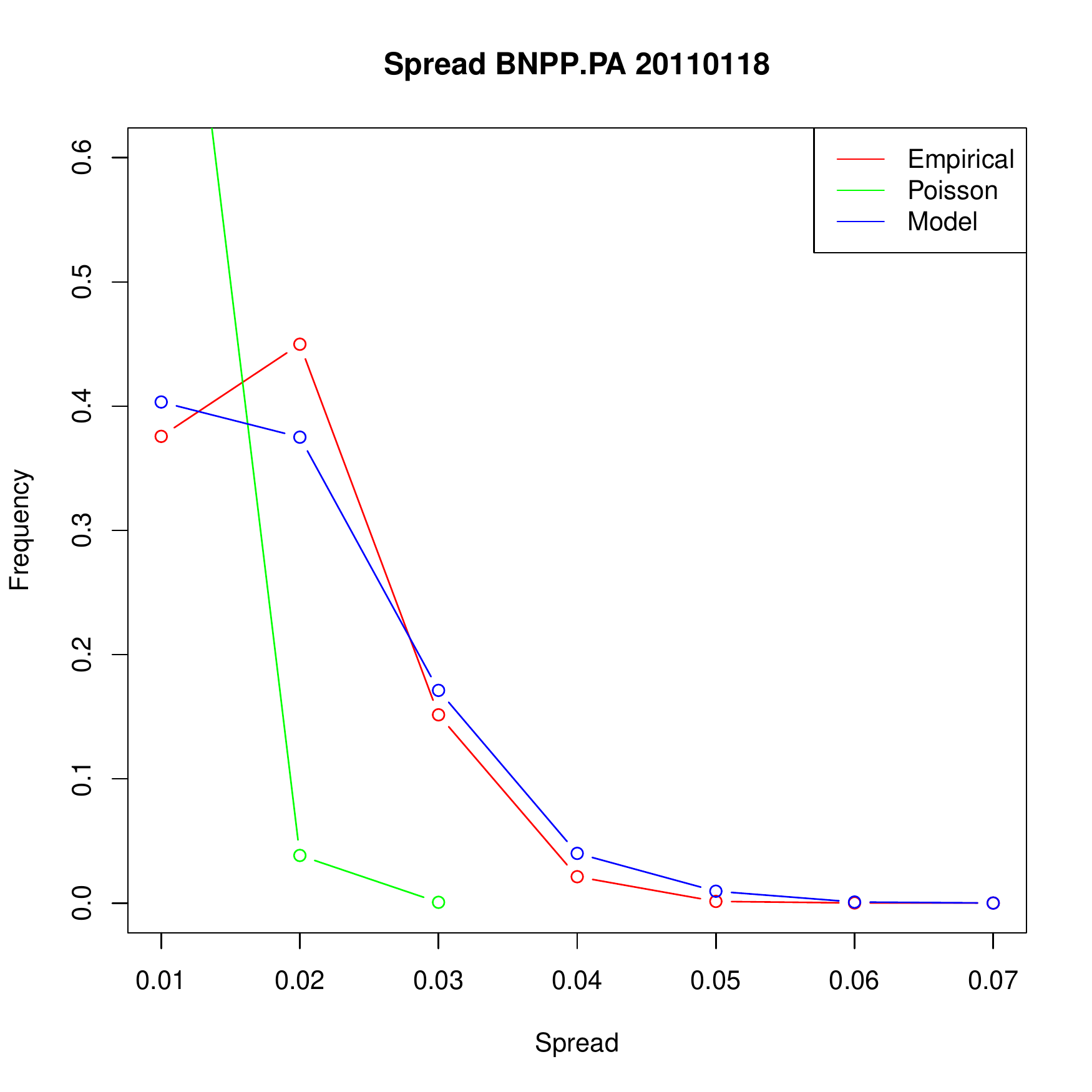}
&
\includegraphics[width=0.4\textwidth, page=1]{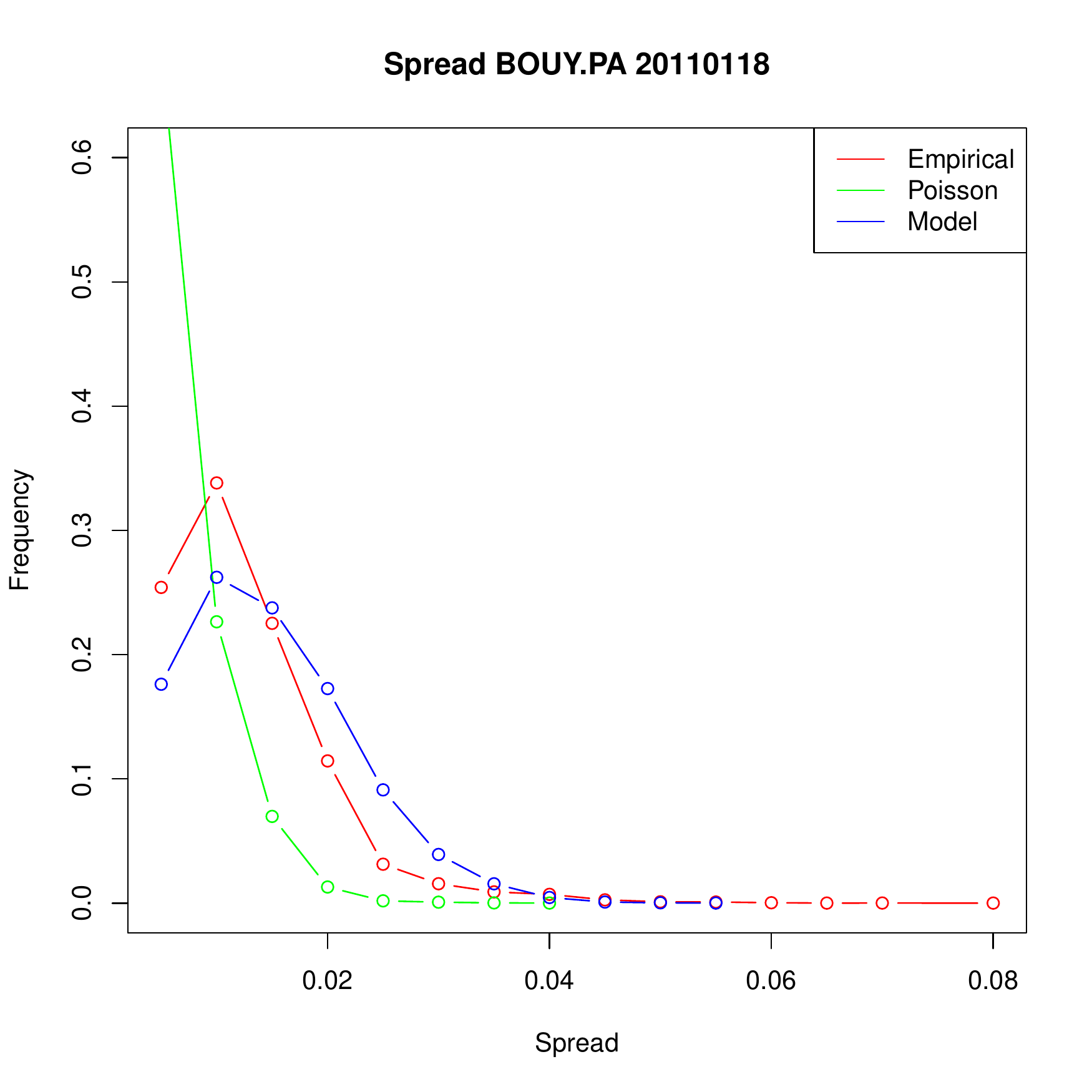}
\\
\includegraphics[width=0.4\textwidth, page=1]{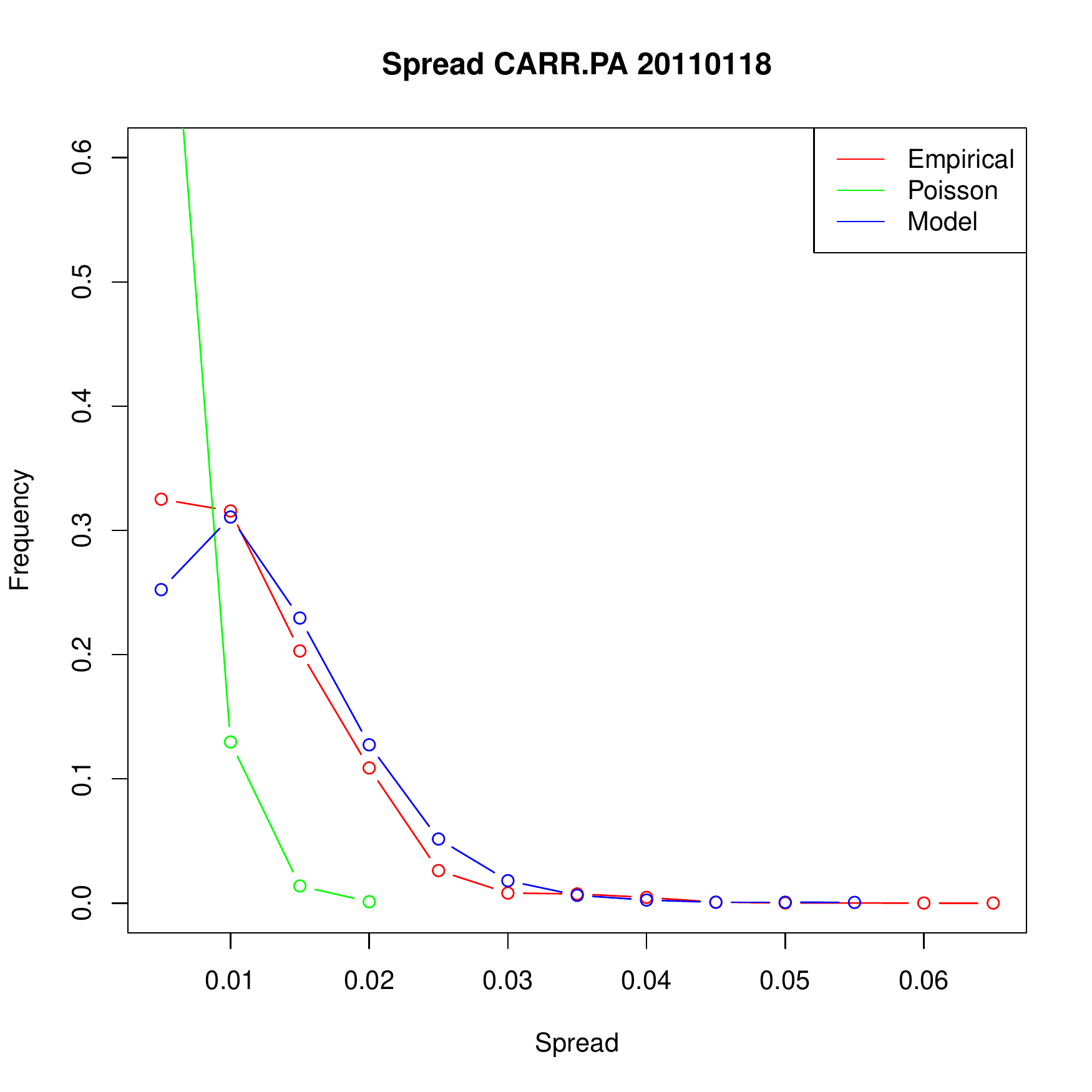}
&
\includegraphics[width=0.4\textwidth, page=1]{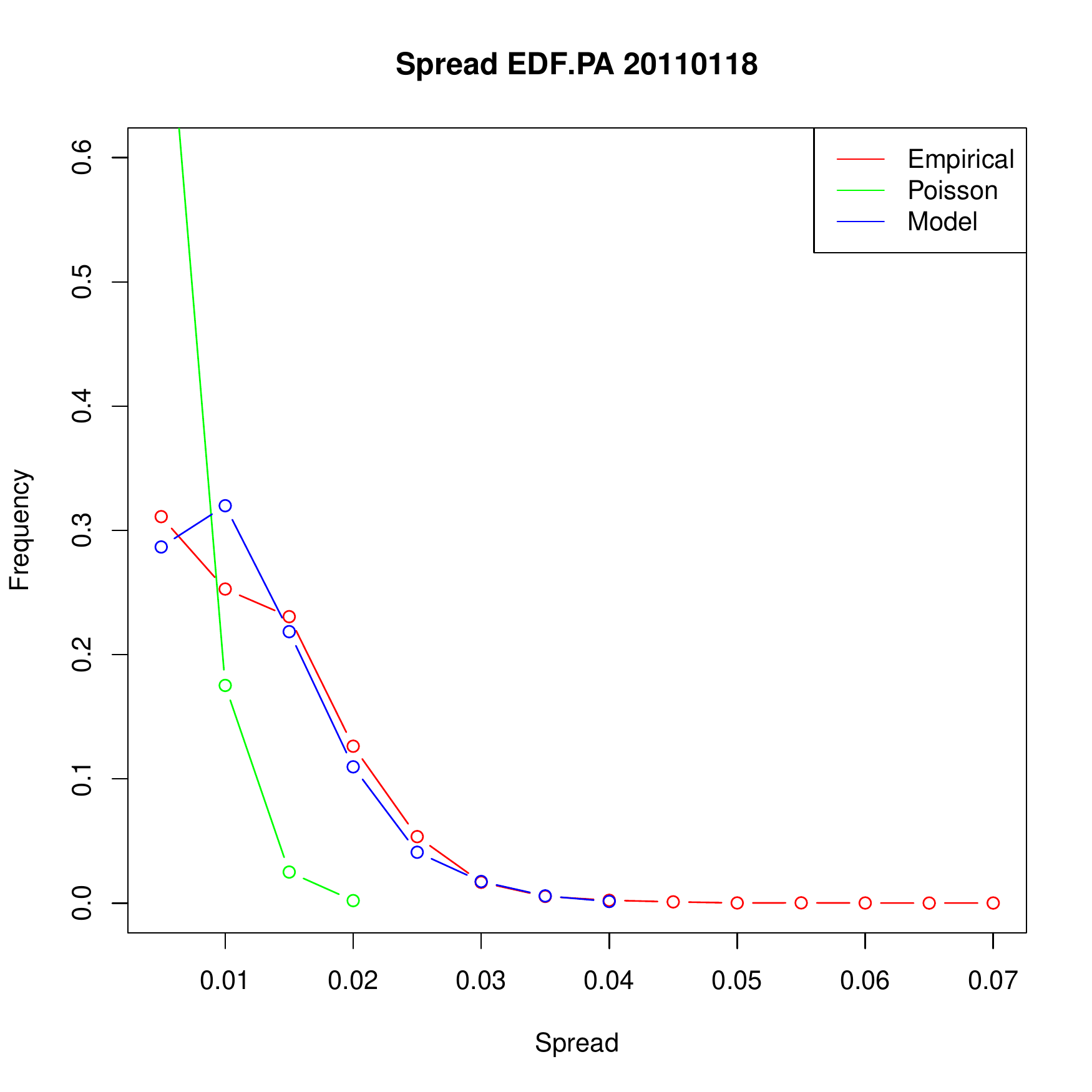}
\end{tabular}
\end{center}
\caption{Distribution of the spread in the model, compared to the empirical distribution and the one produced by a Poisson model. Data: January 18th, 20011.}
\label{figure:Simulator-Spread}
\end{figure}

One of the most important feature is that our model is able to reproduce very well the empirical distribution of the spread. On Figure \ref{figure:Simulator-Spread}, the simulated distribution is a good fit of the empirical one, while the Poisson reference is not relevant at all.
The spread in the Poisson model is most of the time equal to 1 tick, i.e. the book is "stuck". Our model of intensities is able to tackle this problem by increasing the market intensity and decreasing the limit intensity when the spread is small, as it is empirically observed.
It is remarkable to observe that this close fit is obtained for all stocks and dates tested, irrespective of the liquidity and ticksize of the stock studied.

\begin{figure}
\begin{center}
\begin{tabular}{cc}
\includegraphics[width=0.4\textwidth, page=1]{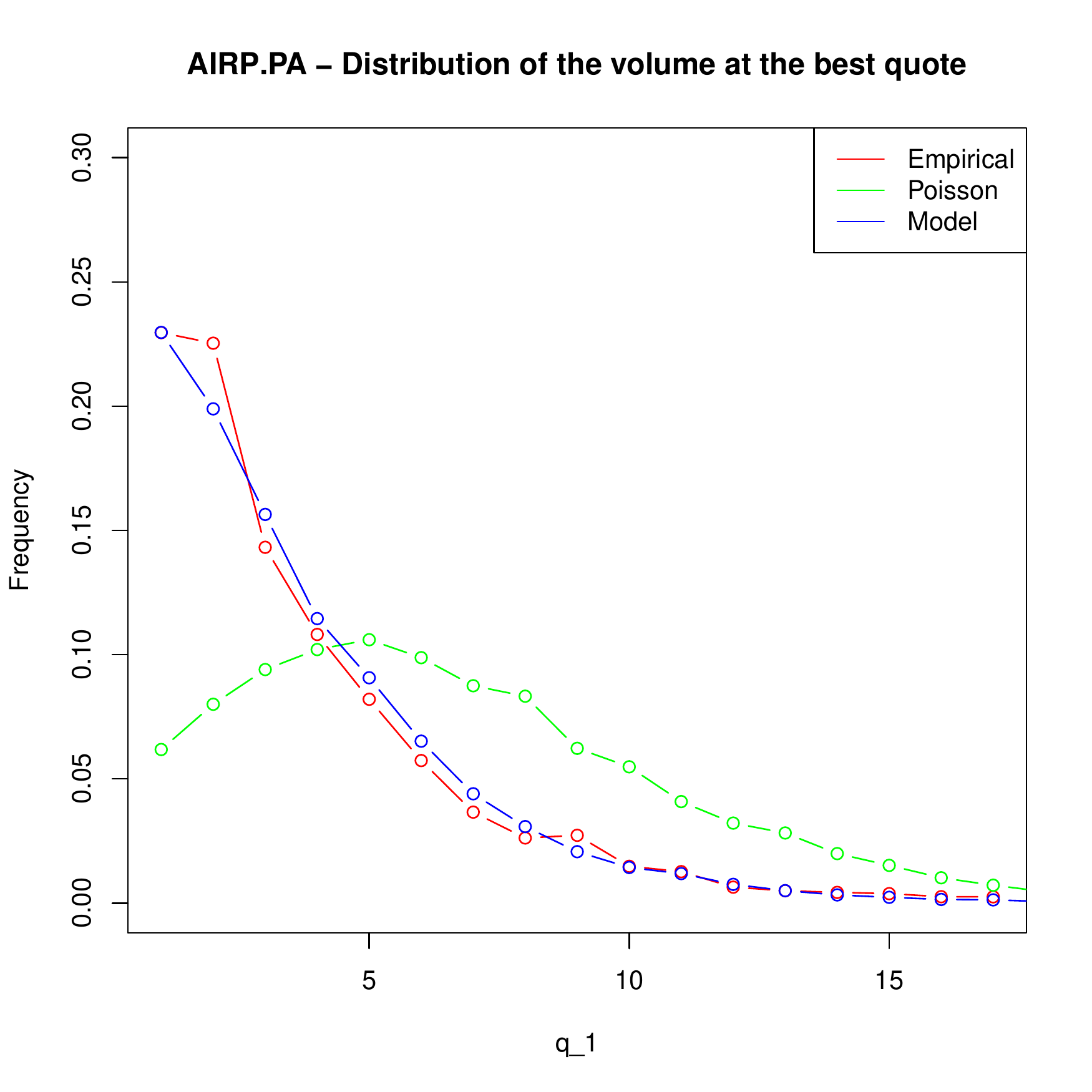}
&
\includegraphics[width=0.4\textwidth, page=1]{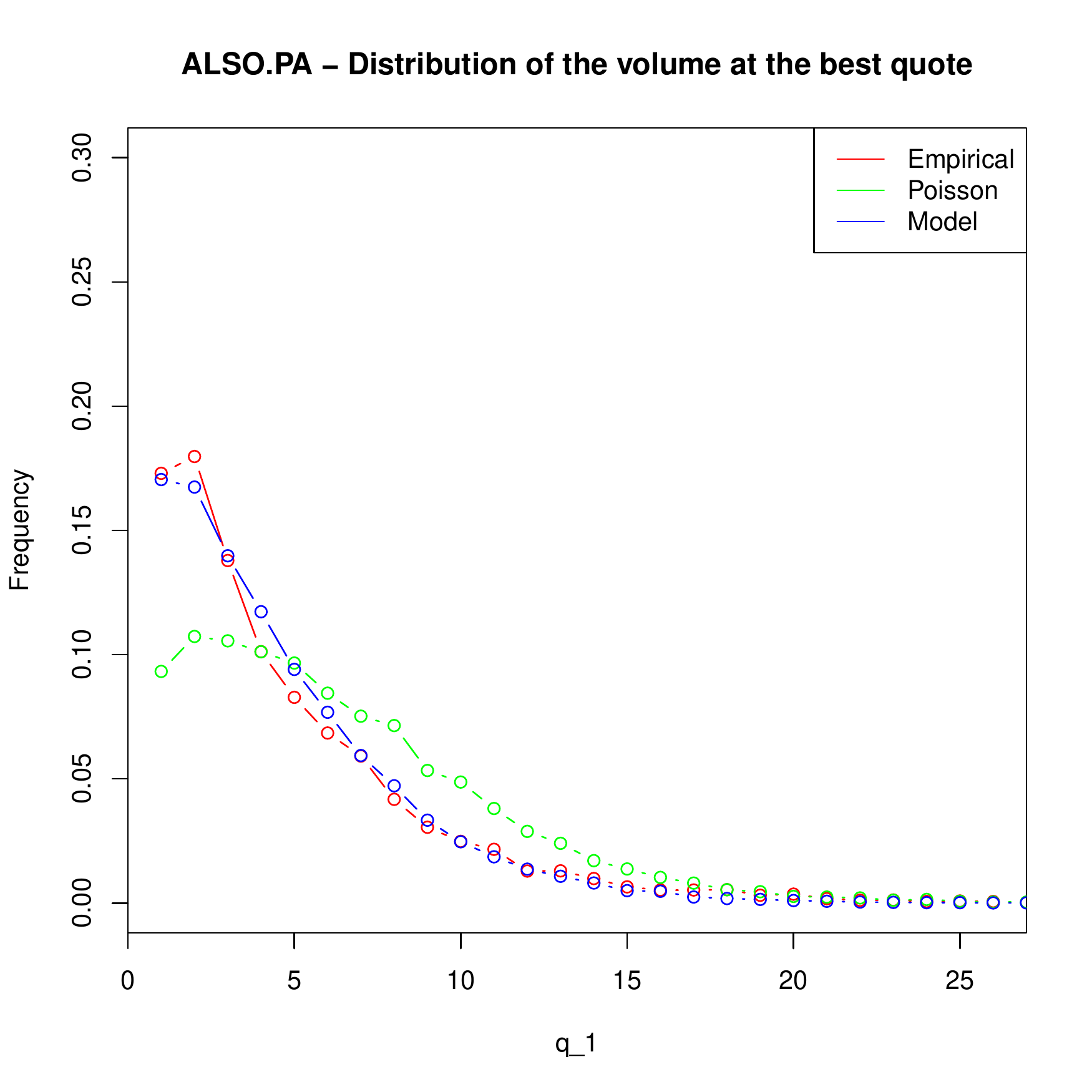}
\\
\includegraphics[width=0.4\textwidth, page=1]{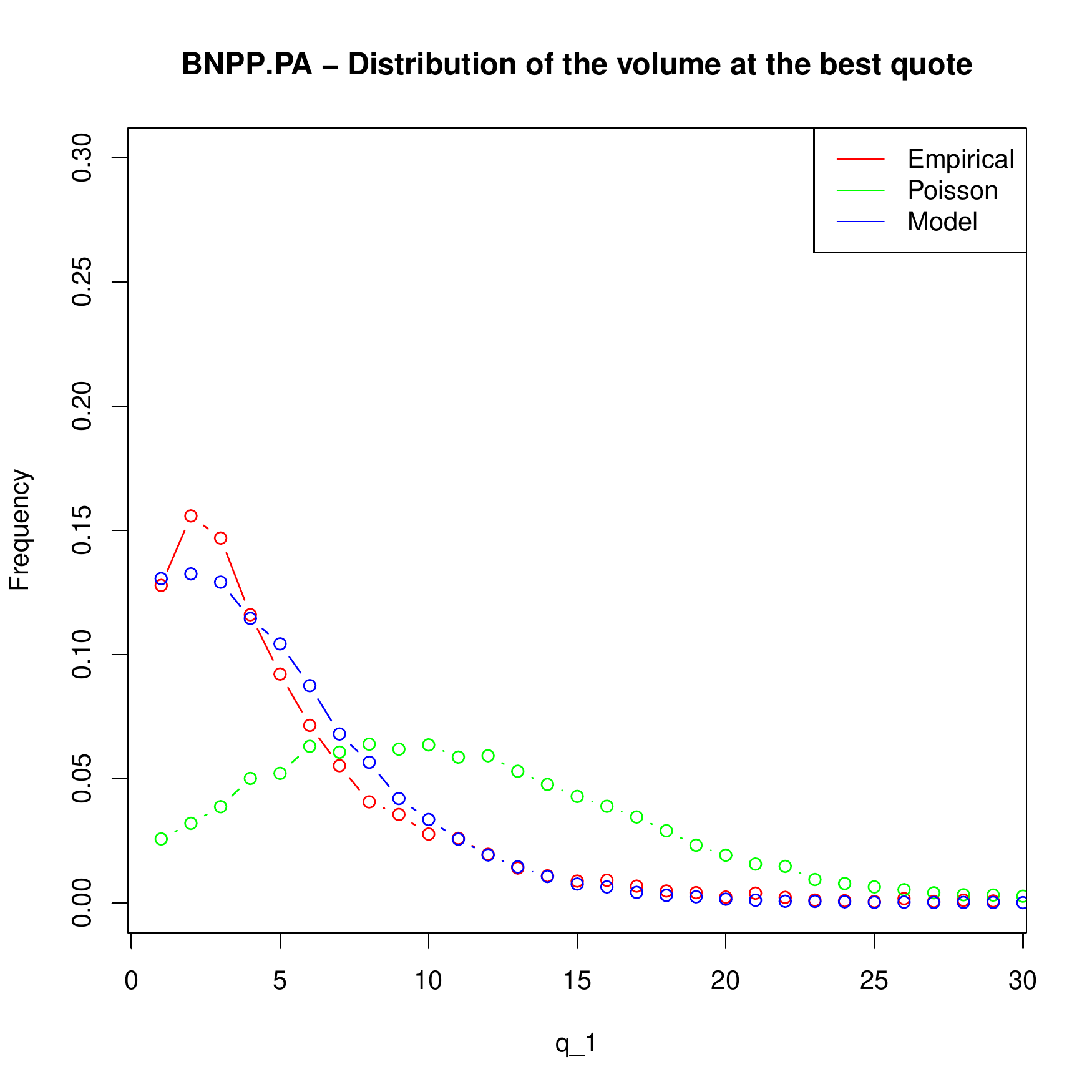}
&
\includegraphics[width=0.4\textwidth, page=1]{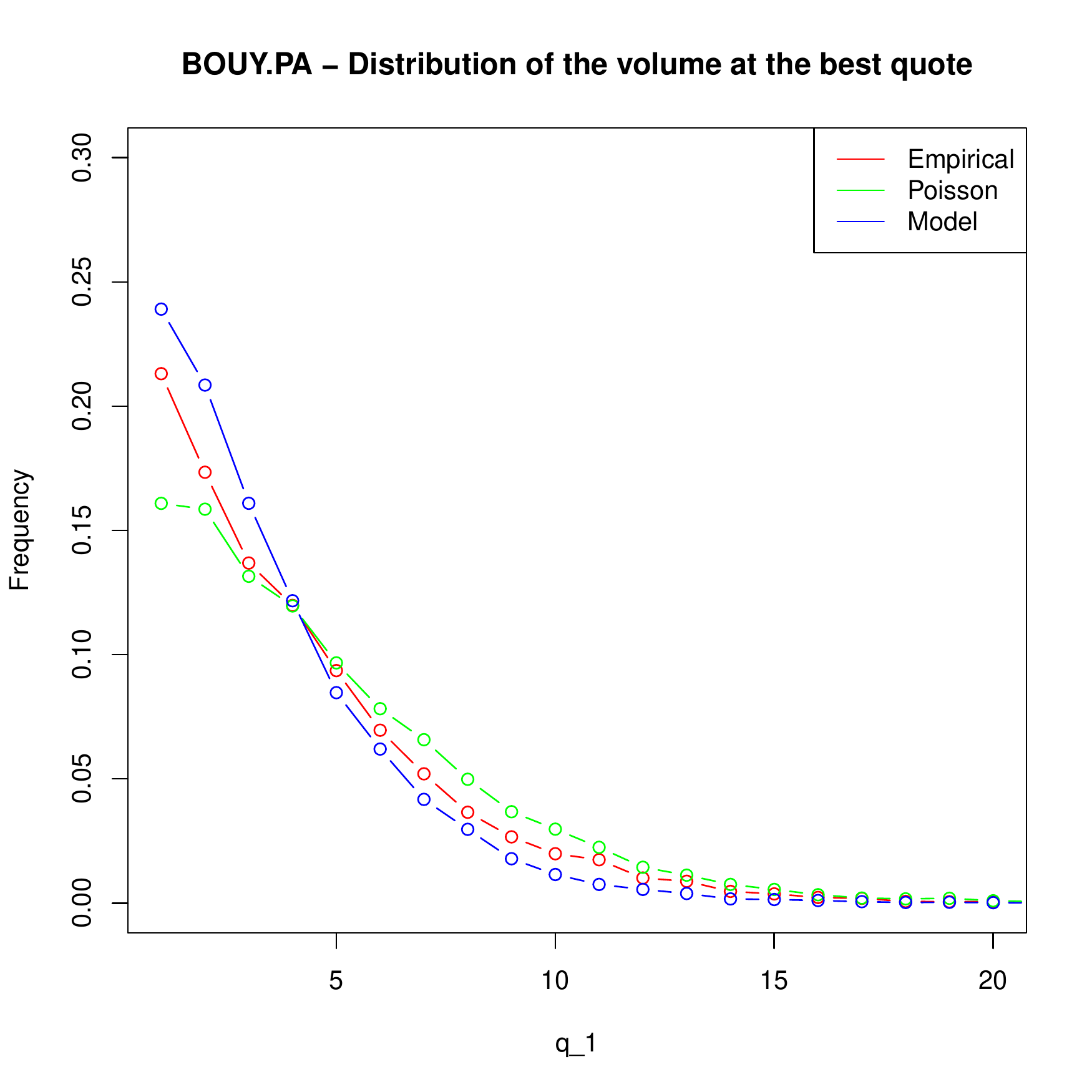}
\\
\includegraphics[width=0.4\textwidth, page=1]{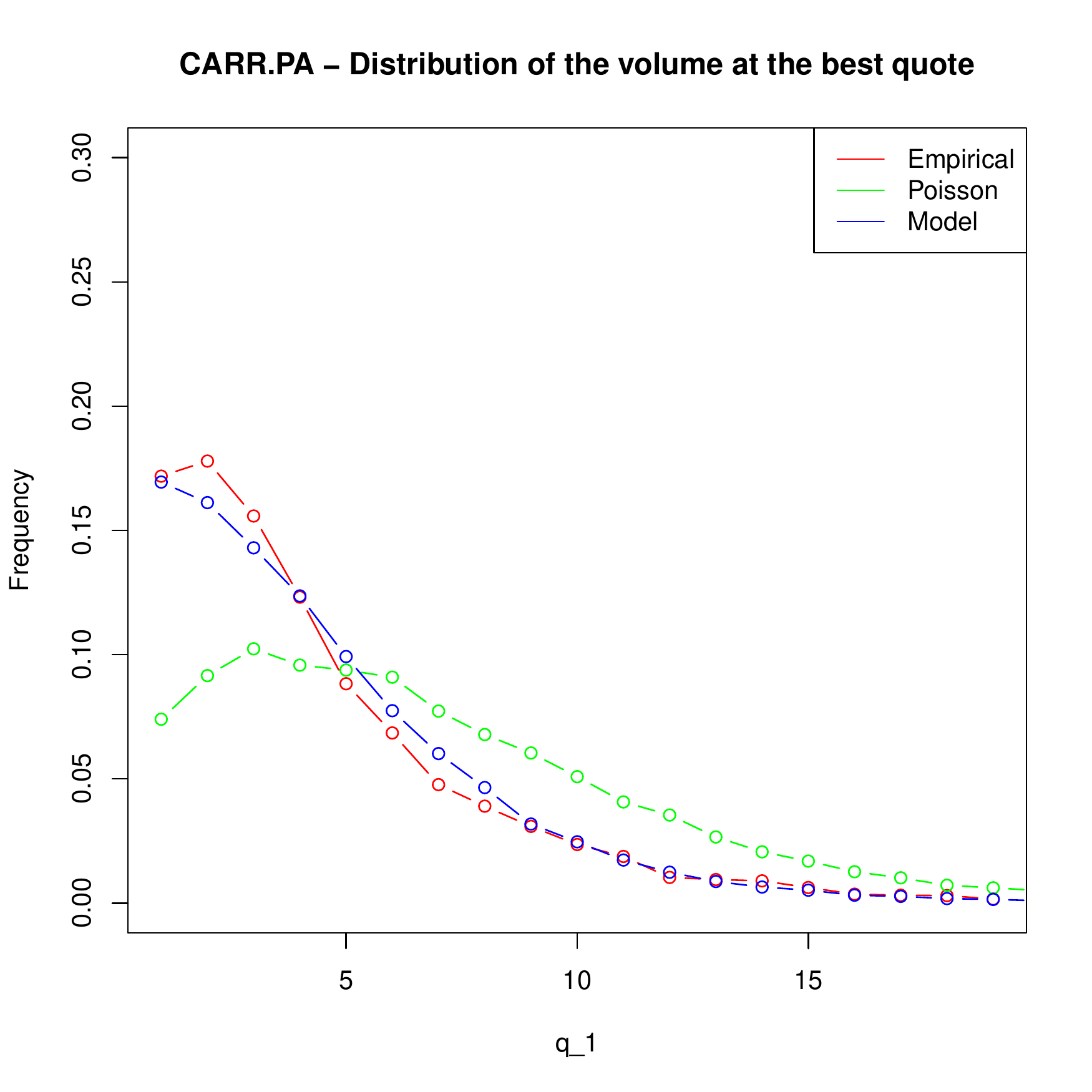}
&
\includegraphics[width=0.4\textwidth, page=1]{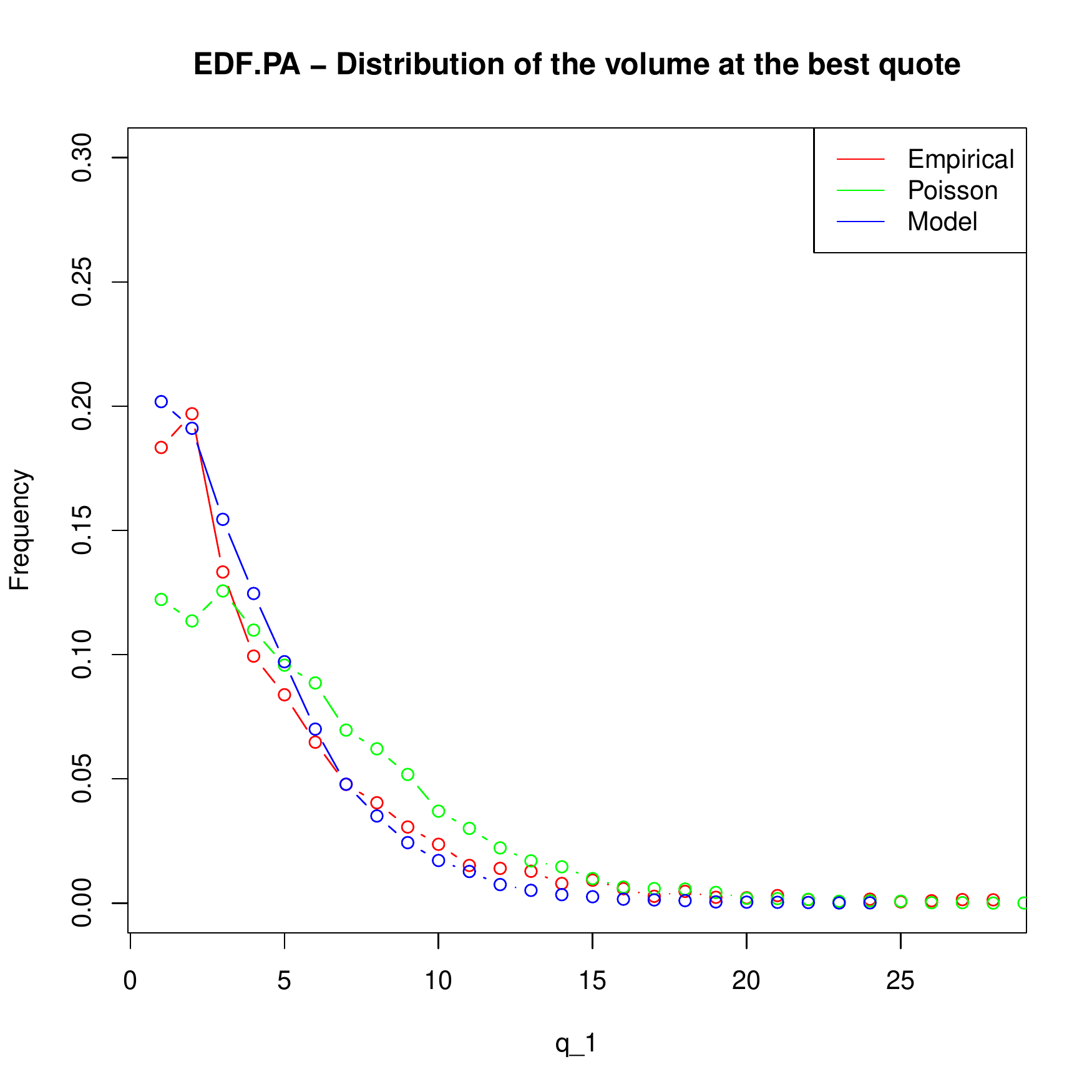}
\end{tabular}
\end{center}
\caption{Distribution of $q_1$ in the model, compared to the empirical distribution and the one produced by a Poisson model. Data: January 18th, 20011.}
\label{figure:Simulator-Level1}
\end{figure}

We now turn to the second modelling variable of our model.
On Figure \ref{figure:Simulator-Level1}, we plot the empirical distribution of $q_1$ and its simulated counterparts.
There again, our model provides an excellent fit for this distribution while the standard Poisson reference constantly underestimates the probability to observe smaller values of $q_1$, i.e. its $q_1$ distribution is shifted to the right. Results are similar for all stocks and dates tested.

\begin{figure}
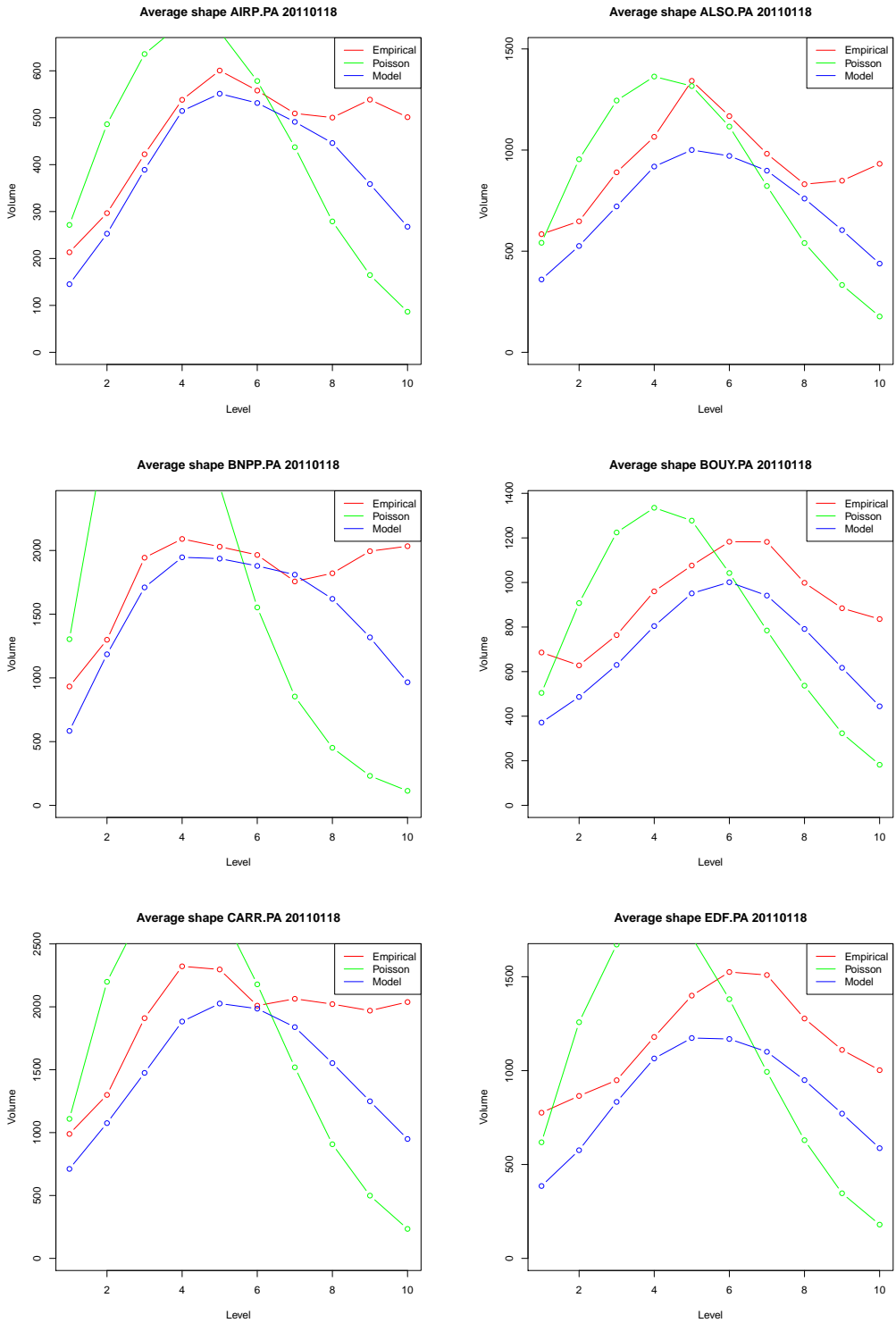

\begin{center}
\begin{tabular}{cc}
\includegraphics[width=0.4\textwidth, page=2]{{AIRP.PA-20110118-CheckAverageShape}.pdf}
&
\includegraphics[width=0.4\textwidth, page=2]{{ALSO.PA-20110118-CheckAverageShape}.pdf}
\\
\includegraphics[width=0.4\textwidth, page=2]{{BNPP.PA-20110118-CheckAverageShape}.pdf}
&
\includegraphics[width=0.4\textwidth, page=2]{{BOUY.PA-20110118-CheckAverageShape}.pdf}
\\
\includegraphics[width=0.4\textwidth, page=2]{{CARR.PA-20110118-CheckAverageShape}.pdf}
&
\includegraphics[width=0.4\textwidth, page=2]{{EDF.PA-20110118-CheckAverageShape}.pdf}
\end{tabular}
\end{center}
\caption{Average shape of the order book in the model, compared to the empirical shape and the one produced by the Poisson model.  Data: January 18th, 20011.}
\label{figure:Simulator-AverageShape}
\end{figure}

If we finally turn to the last variable used in our model, the total volume available $Q_{10}$, then our model is able to reproduce the time average of this quantity. Figure \ref{figure:Simulator-AverageShape} plots the empirical average shape of the order book and the ones produced by the simulators.
Both models are able to quite well reproduce the order of magnitude of average shape of the limit order book. This is not surprising, since the magnitude of the average is directly linked to the way we estimate the parameter $\theta$ in Section \ref{section:Cancellations}, which is identical in both models.
However only our model correctly reproduces the slope of the average order book for the best prices, as well as a sound estimation of the position of the maximum away from the best quotes.
The Poisson reference exhibits a sharper slope for the best prices, realizes a maximum too high and too close to the best quote, and underestimates the volume available far away from the best quotes.
Once again, these observations are valid for all stocks and dates tested.

\begin{figure}
\begin{center}
\begin{tabular}{cc}
\includegraphics[width=0.4\textwidth, page=1]{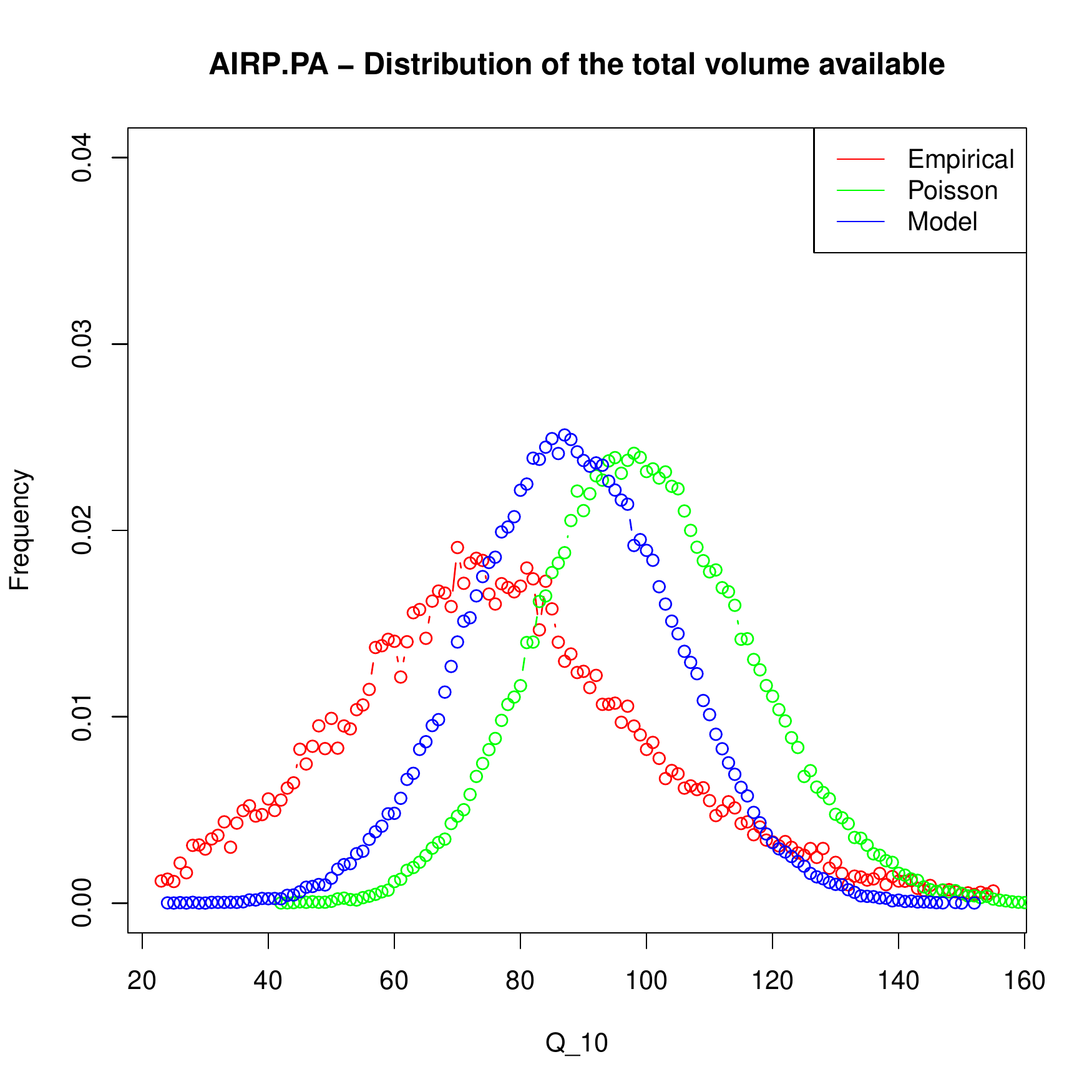}
&
\includegraphics[width=0.4\textwidth, page=1]{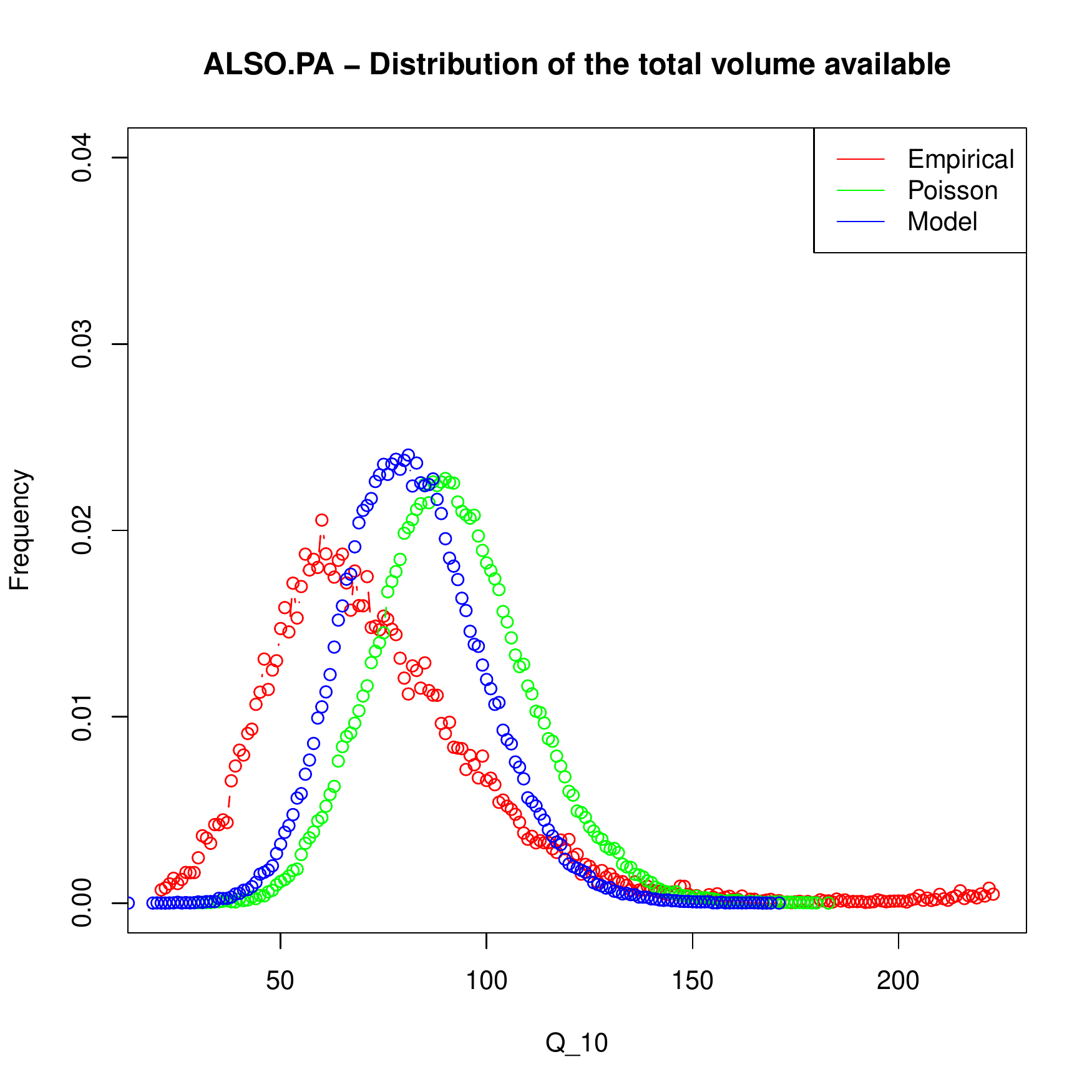}
\\
\includegraphics[width=0.4\textwidth, page=1]{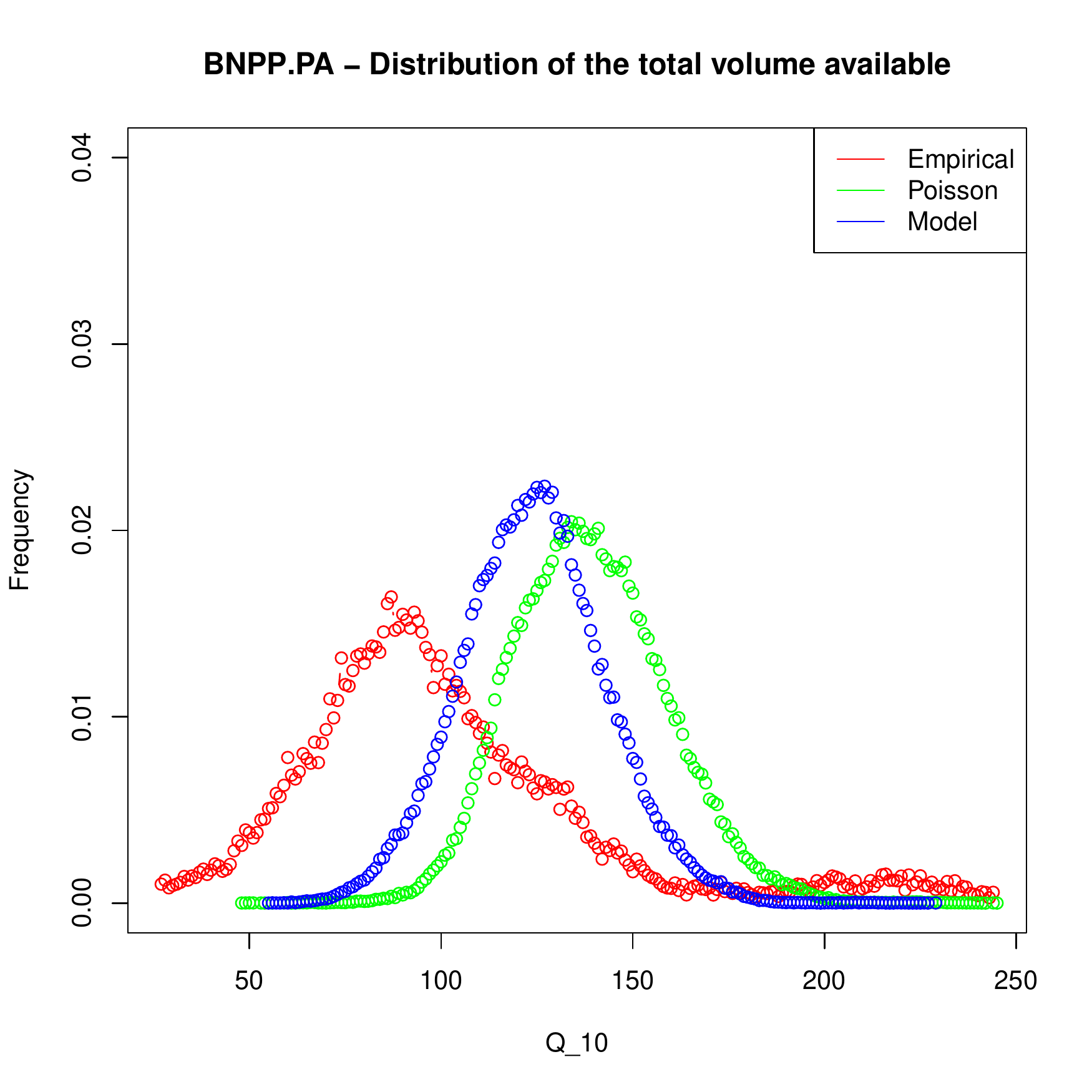}
&
\includegraphics[width=0.4\textwidth, page=1]{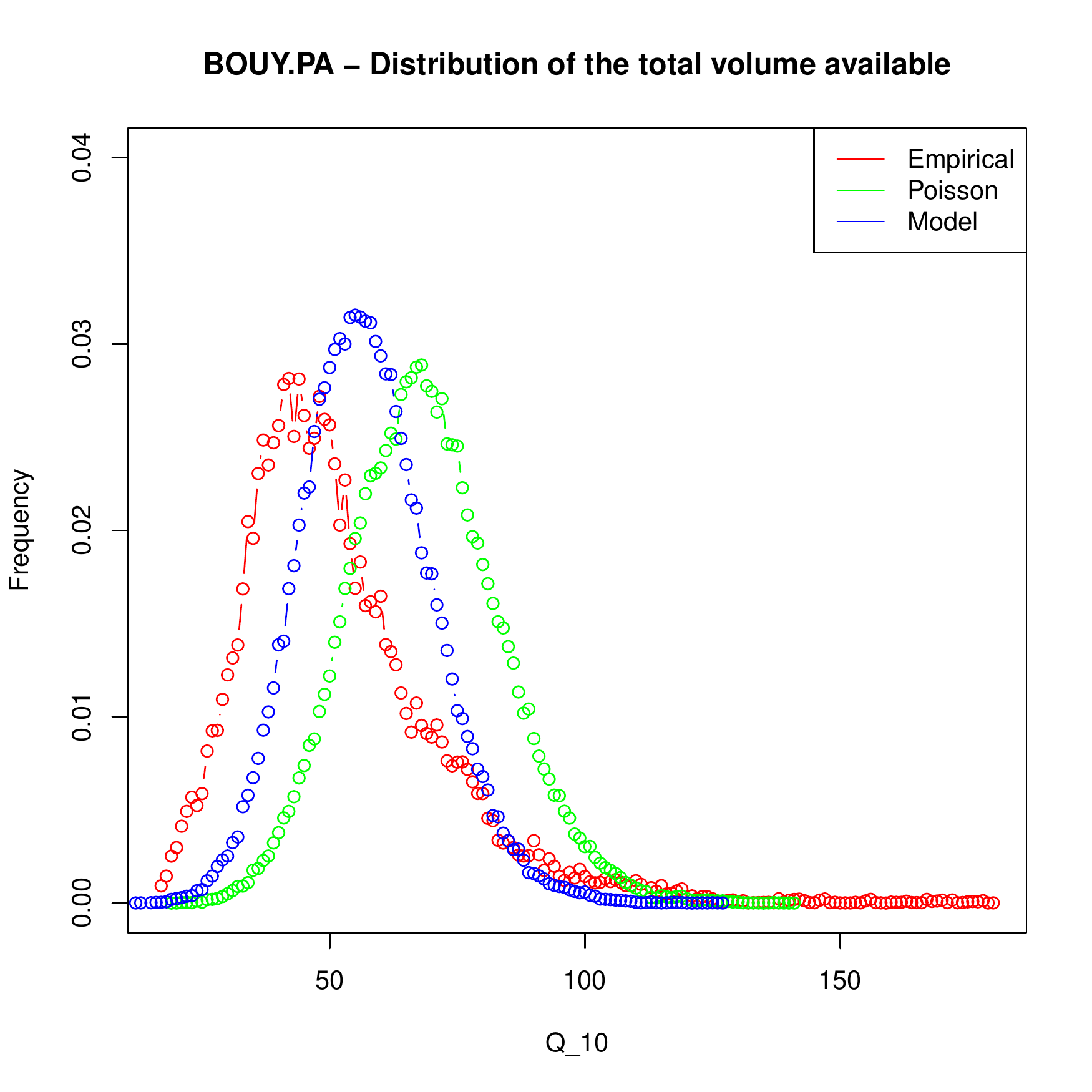}
\\
\includegraphics[width=0.4\textwidth, page=1]{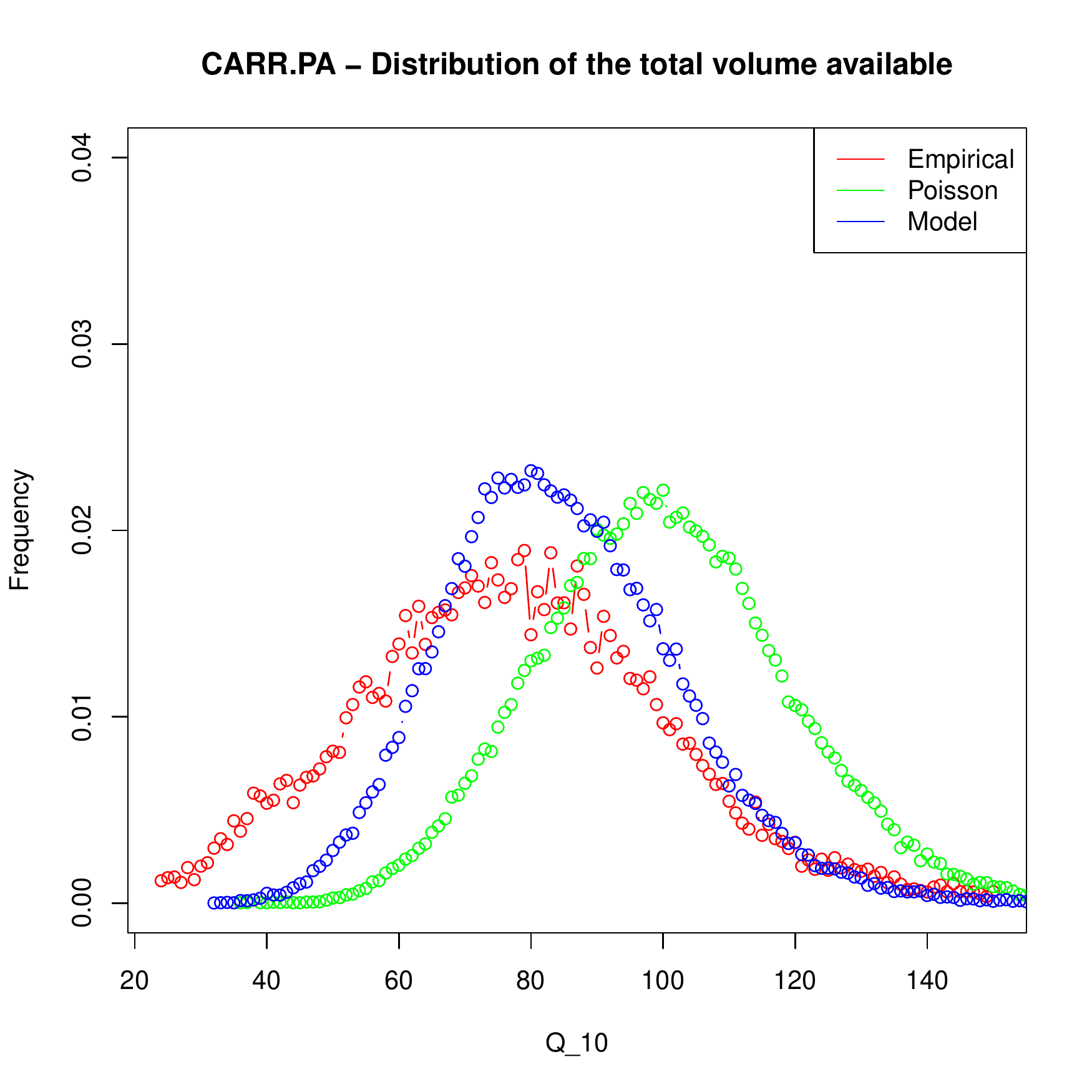}
&
\includegraphics[width=0.4\textwidth, page=1]{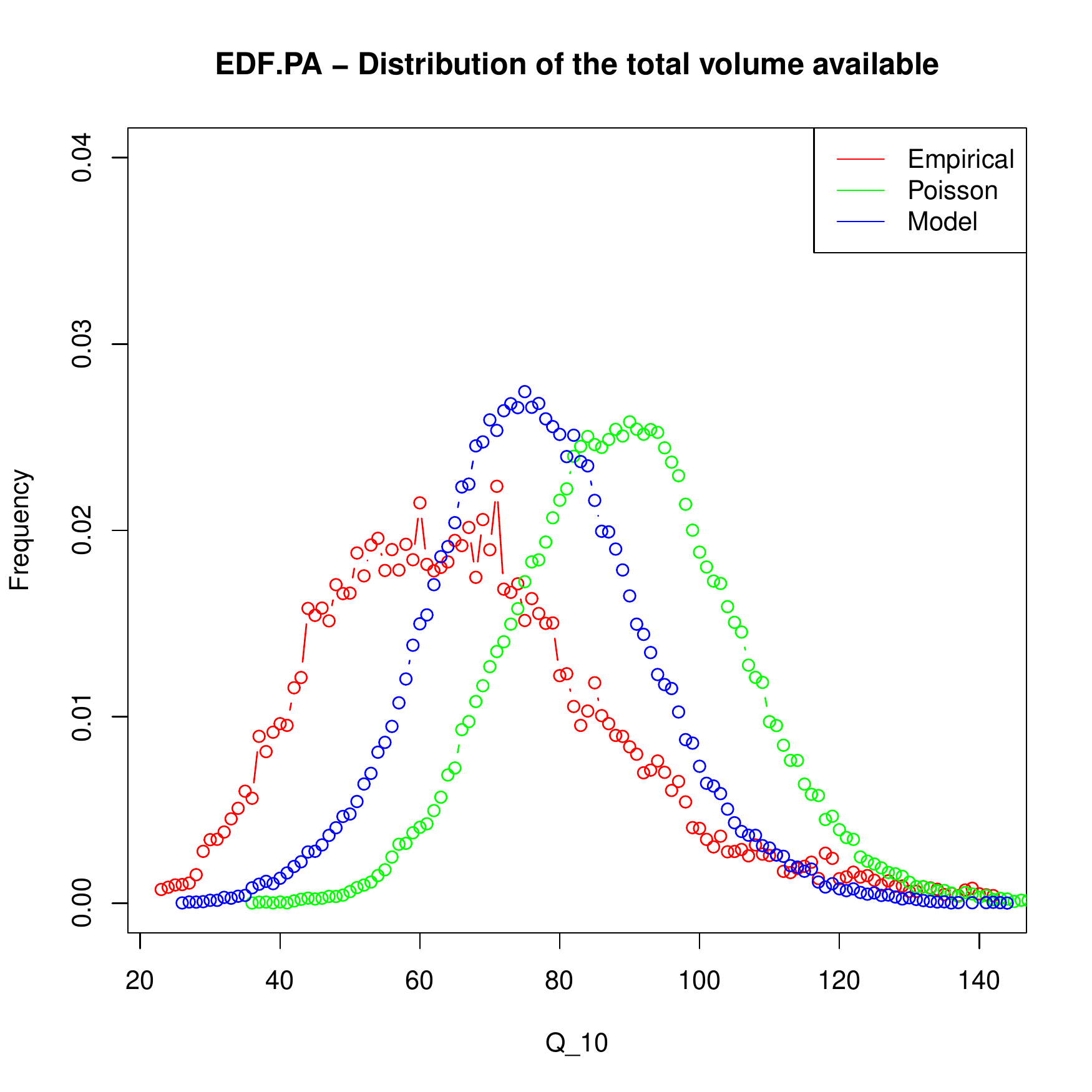}
\end{tabular}
\end{center}
\caption{Distribution of $Q_{10}$ in the model, compared to the empirical shape and the one produced by the Poisson model.}
\label{figure:Simulator-Q10}
\end{figure}

If we go into more details, Figure \ref{figure:Simulator-Q10} plots the empirical distribution of $Q_{10}$.
It turns out that the empirical $Q_{10}$ distribution exhibits a quite heavy tail for large values of $Q_{10}$. Since both models are fitted on the mean, this leads to an underestimation of the probability of lower values of $Q_{10}$ in both simulations. However, the full model outperforms the Poisson reference even in this case.

\section{Conclusion}

We have provided a fully parametric model for the limit order book. The submission of orders is modelled as a point processes with state-dependent intensities.
We provide detailed functional forms for these intensities, as well as the estimation procedure by likelihood maximization.
By developing a market simulator we are able to show that the model performs very well to reproduce key features of the order book, such as the spread and the volume of the best quote in the order book.

This very empirical and numerical work will hopefully lead to further improvements.
The intensities we have proposed here are chosen with respect to some model principles in the choice of variables and functional forms. One may probably go further in the statistical model by experimenting other forms or variables.

This work could also stimulate research on the stability of such complex random systems. Although the mathematics of the "Poisson" models for the order book are beginning to be well-understood, the introduction of state-dependent intensities could lead to several theoretical problems that have not been studied here.

\bibliographystyle{agsm}
\bibliography{ModellingIntensities}

\end{document}